\documentclass[pdflatex,sn-mathphys-num,iicol]{sn-jnl}

\usepackage{graphicx}%
\usepackage{multirow}%
\usepackage{amsmath,amssymb,amsfonts}%
\usepackage{amsthm}%
\usepackage{mathrsfs}%
\usepackage[title]{appendix}%
\usepackage{textcomp}%
\usepackage{manyfoot}%
\usepackage{booktabs}%
\usepackage{algorithm}%
\usepackage{algorithmicx}%
\usepackage{algpseudocode}%
\usepackage{listings}%

\usepackage[]{rotating,times,pdfsync,epstopdf,amsfonts}
\usepackage{mhchem}
\usepackage{amsmath}
\usepackage{amssymb}
\usepackage{empheq}
\usepackage{color}

\usepackage{bm}
\usepackage{gensymb}
\usepackage{upgreek}
\usepackage{caption}
\usepackage{subcaption}
\usepackage{hhline}
\usepackage[table]{xcolor}

\begin{document}

\title[Article Title]{Antiproton annihilation at rest in thin solid targets and comparison with Monte Carlo simulations}

\author[1]{\fnm{C.} \sur{Amsler}} 
\author[2]{\fnm{H.} \sur{Breuker}}  
\author[3,4]{\fnm{M.} \sur{Bumbar}}
\author[1,4]{\fnm{M.} \sur{Cerwenka}} 
\author[5,6]{\fnm{G.} \sur{Costantini}} 
\author[7,8]{\fnm{R.} \sur{Ferragut}} 
\author[1,a]{\fnm{M.} \sur{Fleck}} 
\author[8]{\fnm{M.} \sur{Giammarchi}} 
\author*[1,b]{\fnm{A.} \sur{Gligorova}}\email{angela.gligorova@univie.ac.at}  
\author[5,6]{\fnm{G.} \sur{Gosta}}

\author[1,c]{\fnm{E. D.} \sur{Hunter}} 
\author[1]{\fnm{C.} \sur{Killian}}
\author[1,d]{\fnm{B.} \sur{Kolbinger}}
\author[1,4]{\fnm{V.} \sur{Kraxberger}} 
\author[9]{\fnm{N.} \sur{Kuroda}} 
\author[1]{\fnm{M.} \sur{Lackner}} 

\author[5,6]{\fnm{M.} \sur{Leali}}
\author[8,10]{\fnm{G.} \sur{Maero}} 
\author[5,6]{\fnm{V.} \sur{Mascagna}}
\author[9]{\fnm{Y.} \sur{Matsuda}}  
\author[5,6]{\fnm{S.} \sur{Migliorati}} 
\author[1]{\fnm{D. J.} \sur{Murtagh}} 
\author[1,4]{\fnm{A.} \sur{Nanda}}  
\author[3,4]{\fnm{L.} \sur{Nowak}} 
\author[1]{\fnm{S.} \sur{Rheinfrank}}
\author[8,10]{\fnm{M.} \sur{Romé}} 
\author[1]{\fnm{M. C.} \sur{Simon}} 
\author[11,e]{\fnm{M.} \sur{Tajima}}  
\author[8,10]{\fnm{V.} \sur{Toso}} 
\author[2]{\fnm{S.} \sur{Ulmer}}  

\author[12]{\fnm{M.} \sur{van~Beuzekom}}
\author[5,6]{\fnm{L.} \sur{Venturelli}}
\author[1,4]{\fnm{A.} \sur{Weiser}} 
\author[1]{\fnm{E.} \sur{Widmann}} 
\author[2]{\fnm{Y.} \sur{Yamazaki}}  
\author[1]{\fnm{J.} \sur{Zmeskal}} 

\affil[1]{\orgname{Stefan Meyer Institute for Subatomic Physics, Austrian Academy of Sciences}, \orgaddress{ \city{Vienna},  \country{Austria}}} 
\affil[2]{\orgname{Ulmer Fundamental Symmetries Laboratory, RIKEN} \orgaddress{\city{Saitama}, \country{Japan}}} 
\affil[3]{\orgname{Experimental Physics Department, CERN}, \orgaddress{\city{Geneva}, \country{Switzerland}}} 
\affil[4]{\orgname{University of Vienna, Vienna Doctoral School in Physics}, \orgaddress{\city{Vienna}, \country{Austria}}} 
\affil[5]{\orgname{Dipartimento di Ingegneria dell’Informazione, Università degli Studi di Brescia}, \orgaddress{\city{Brescia}, \country{Italy}}}
\affil[6]{\orgname{INFN sez. Pavia}, \orgaddress{\city{Pavia}, \country{Italy}}} 
\affil[7]{\orgname{L-NESS and Department of Physics, Politecnico di Milano}, \orgaddress{\city{Como}, \country{Italy}}} 
\affil[8]{\orgname{INFN sez. Milano}, \orgaddress{\city{Milan}, \country{Italy}}} 

\affil[9]{\orgname{Institute of Physics, Graduate School of Arts and Sciences, University of Tokyo}, \orgaddress{\city{Tokyo}, \country{Japan}}} 
\affil[10]{\orgname{Dipartimento di Fisica, Università degli Studi di Milano}, \orgaddress{\city{Milan}, \country{Italy}}}

\affil[11]{\orgname{RIKEN Nishina Center for Accelerator-Based Science}, \orgaddress{\city{Saitama}, \country{Japan}}} 
\affil[12]{\orgname{Nikhef National Institute for Subatomic Physics}, \orgaddress{\city{Amsterdam}, \country{Netherlands}}} 
\affil[a] {\orgname{present address: University College London, Department of Physics \& Astronomy}, \orgaddress{\city{London}, \country{United Kingdom}}} 
\affil[b] {\orgname{present address: University of Vienna, Faculty of Physics}, \orgaddress{\city{Vienna}, \country{Austria}}}
\affil[c] {\orgname{present address: Experimental Physics Department, CERN}, \orgaddress{\city{Geneva}, \country{Switzerland}}}
\affil[d]{\orgname{present address: DATA TECHNOLOGY},
\orgaddress{\city{Vienna}, \country{Austria}}}
\affil[e]{\orgname{present address: Japan Synchrotron Radiation Research Institute}, \orgaddress{\city{Hyogo}, \country{Japan}}}


\abstract{The mechanism of antiproton-nucleus annihilation at rest is not fully understood, despite substantial previous experimental and theoretical work. In this study we used slow extracted antiprotons from the ASACUSA apparatus at CERN to measure the charged particle multiplicities and their energy deposits from antiproton annihilations at rest on three different nuclei: carbon, molybdenum and gold. The results are compared with predictions from different models in the simulation tools Geant4 and FLUKA.
A model that accounts for all the observed features is still missing, as well as measurements at low energies, to validate such models.}


\maketitle

\section{\label{sec:intro}Introduction}

Antiproton ($\mathrm{\bar{p}}$) annihilation on a nucleus (A) is a particular nuclear reaction in which part of the released energy is transferred to the nucleus with reduced linear and angular momentum, compared to collisions with protons or heavy ions. Consequently, the decay mechanism of nuclei heated by stopped and energetic antiprotons allows the observation of some effects characteristic of heavy-ion reactions, such as multifragment disintegration, without nuclear compression~\cite{Plendl_1993, gold_fragm, PS203}. Our understanding of the mechanism of antiproton annihilation in nuclear matter is based on observables such as multiplicity, energy and angular distributions of secondary particles. Various models, including the Intranuclear Cascade (INC) model~\cite{INC_1, INC_ILJINOV82}, the statistical multifragmentation model~\cite{stat_multifrag_model} or the Lanzhou quantum molecular-dynamics approach~\cite{Lanzhou} describe the nuclear dynamics and decay mechanism induced by antiprotons with varying degrees of success. A great amount of complexity in the antiproton-nucleus annihilation comes from the rich variety of possible reaction channels triggered by the primary annihilation mesons.  

In addition, the antiproton annihilation process is still not fully understood at a quark level and the microscopic models usually include phenomenological parameters that have been tuned to reproduce early and incomplete sets of data. The quark rearrangement model explains the correct order of magnitude for the branching ratios of $\mathrm{\bar{p}}$p annihilation at rest into three mesons. However, to account for two-body modes, kaon production and more precise predictions of branching ratios, other contributions, such as from quark-antiquark annihilation have to be included. A phenomenology was developed to try to extract from the data the relative importance of the various types of diagrams, but there is no consensus as to which model prevails~\cite{pbar_report_2005}. 

Annihilation at rest most of the time occurs on a single nucleon in the peripheral nuclear region, where the nucleon density is less than 10\% of the central density~\cite{INC_ILJINOV82}. For $\mathrm{\bar{p}p}$ annihilation, the total energy of 1877~MeV is converted into an average of 3.0$\pm$0.2 charged pions ($\pi^{\pm}$) and 2.0$\pm$0.2 neutral pions ($\pi^{0}$), with an average kinetic energy of 230~MeV~\cite{claude1998,pbar_report_2005}. Kaons and $\eta$ mesons are produced in about 6\% and 7\% of all annihilations, respectively. When the annihilation occurs on a neutron, on average 1.07$\pm$0.04~$\pi^{+}$ and 2.07$\pm$0.04~$\pi^{-}$ are produced~\cite{Bendiscioli1990}. 
Absorption or scattering (final state interactions, FSI) of the emitted mesons may follow, with or without break-up of the nucleus, which depends upon the solid angle under which the remaining nucleons are seen from the annihilation point~\cite{Cugnon2001_geo_eff}, and which may modify the detected meson distribution~\cite{FSI1}. While many annihilation channels have been measured for antiproton-nucleon ($\mathrm{\bar{p}}$N) annihilation at rest, for $\mathrm{\bar{p}}$A, the branching ratios have been hitherto reported only for $\mathrm{^{3}He}$ and $\mathrm{^{4}He}$~\cite{topics_pabr_nucleus,Bendiscioli1990, 4He_annihilation}, and the total prong (charged particle) multiplicity has been measured for a very limited number of nuclei~\cite{Bendiscioli}. Furthermore, even for $\mathrm{\bar{p}N}$ annihilations the current models do not perform satisfactorily in describing all the aspects of the measured data.

The bulk of $\mathrm{\bar{p}A}$ data originates from LEAR (Low Energy Antiproton Ring) experiments, taken about three decades ago by degrading $\sim$20~MeV antiprotons in a number of moderators, and ultimately stopping them in thick targets. In these measurements, only light prongs were detected (helium and lithium ions being the heaviest), in certain energy intervals and within narrow solid angles~\cite{PS203, spectra_pbar, Plendl_1993, pbarA_helium_ions}. Features such as particle yields and momentum spectra of the emitted pions, protons (p), deuterons (d) and tritons (t) with energies above 10~MeV were reported, as well as the energy dissipation (energy transfer to the nucleons and mean excitation energies \cite{PS203}). The residual nuclei have been investigated through measurements of their radioactivity after $\mathrm{\bar{p}}$ irradiation, \textit{i.e.} of their gamma spectra~\cite{MOSER198625,vonEgidy1990, Moser:1989ds,PRL_halo, residues_halo, PS209}, and the neutron density distributions were deduced through nuclear spectroscopy analysis of the antiproton annihilation residues~\cite{neutron_density}. The experimental results have been most extensively compared to the INC, which models the $\mathrm{\bar{p}}$A annihilation through the antiproton annihilation on a single nucleon, generating a number of pions which then cascade through the nucleus in a sequence of interactions with the rest of the nucleons. Good agreement with data was found on the absolute pion and proton yields, however the momentum distribution of pions and protons emerging from the annihilation were described with less accuracy~\cite{Cugnon1989}. For the residue distribution, the evaporation model provides a satisfactory description~\cite{INC_review}.

Nevertheless, some characteristics of $\mathrm{\bar{p}}$A annihilation at rest, such as hadronization and total final product multiplicities remain unknown. The production of fragments with short ranges has also not been studied and there is no experimental evidence for heavier nuclei than $\mathrm{^{8}He}$ in the energy distributions~\cite{pbarA_helium_ions}. The charge and mass distribution of the highly ionizing nuclear fragments carry information on the energy deposited during the intranuclear cascade. 

Besides its relevance for nuclear physics studies, the antiproton-nucleus annihilation at rest is one of the key processes for experiments at the Antiproton Decelerator (AD) at CERN, that study antimatter in laboratory, detecting it through annihilation~\cite{asacusa_hodo, asacusa_nagata, aegis_FACT, alpha_SIDET, alpha_TPC, PUMA}. Monte Carlo simulations are performed to assess the efficiency of tagging antihydrogen events~\cite{asacusa_geant4_det}, relying on the physics assumptions implemented either in Geant4, such as the CHiral Invariant Phase Space event generator (CHIPS)~\cite{CHIPS_pbarp, CHIPS_pions, chips_3}, Fritiof-Precompound (FTFP)~\cite{Fritiof1, Fritiof2, FTFP} and the newly added Intranuclear Cascade de Li\`ege (INCL) model~\cite{demid_thesis_incl}, or in FLUKA~\cite{FLUKA_CERN1, FLUKA_CERN2, FLUKA_WEB}. The low energy annihilation models used in these software packages are, however, based on hadronic high energy interactions (CHIPS, FTFP) or were developed for medical physics applications (FLUKA), and were extrapolated to low energies in spite of the fact that the low-energy annihilation mechanism is still not well understood~\cite{amsler2019ect}. Recent studies have been performed using 5~MeV pulsed antiproton beam from the AD, degrading it to energies of $\sim$100~keV in a broad distribution. Antiprotons annihilating at rest in Al, Cu, Ag, Au and Si were measured with emulsion and silicon detectors, as reported in \cite{emulsion2,mimotera, emulsion1}. The results revealed significant discrepancies in the prong multiplicity between measured data and the Geant4 models, with differences ranging from 30\% to a factor of 4.

In this work, we present new measurements of $\mathrm{\bar{p}}$A annihilation at rest using a slow extracted, monoenergetic beam of 150~eV, accelerated to 1.15~keV by the electric potential applied on the solid target. The antiprotons annihilated on thin targets, providing a crucial comparison with the models in which the stopping effects and production of secondaries due to the particle propagation can be neglected. In addition, this allows the emergence and detection of the heavy nuclei produced in the annihilation. The presented work initiates an extended experimental study at ELENA~\cite{ELENA}, along with state-of-the art detector technologies, to study in detail the production of various charged particles in antiproton induced reactions in different nuclei (see Sec.~\ref{sec:conclusions}). The ultimate goal is to provide a deeper understanding of antiproton-nucleus annihilation through quantitative and qualitative analyses of FSIs and their and their dependence on the atomic number.

\section{\label{sec:exp_detail}Experimental setup}

\begin{figure*}[h]
    \centering
    \includegraphics[width=0.8\textwidth]{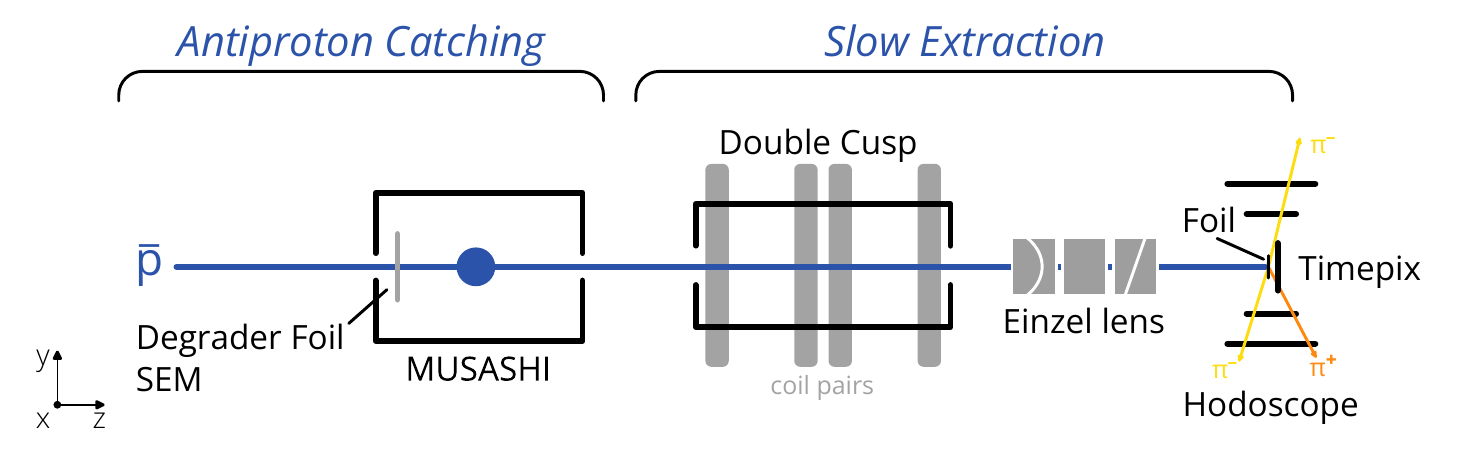}
    \caption{Schematic overview of the experimental setup. The x-axis points to the reader.}
    \label{fig:whole_setup}
\end{figure*}

The ASACUSA-Cusp experiment at CERN, the main purpose of which is to measure the ground-state hyperfine splitting of antihydrogen~\cite{asacusa_web, Widmann_HFS}, is also suited for studies with a continuous (DC) beam of sub-keV~antiprotons. The apparatus was supplied with short pulses of 5.3~MeV antiprotons ($\sim3\cdot~\mathrm{10^{7}}$~$\mathrm{\bar{p}}$/shot, every 100~s) from the AD, decelerated to 120 keV by a radio frequency quadrupole decelerator~\cite{rfqd} and further to $\sim$10~keV with degrader foils.
The slow extraction cycle ($\sim$140~s) started with antiproton trapping in the MUSASHI (Monoenergetic Ultra-Slow Antiproton Source for High-precision Investigation) trap, followed by electron cooling for 40~s without radial compression. The antiprotons were then released by lowering the trapping potential over 30--40~s, as described in ref.~\cite{ASACUSA_DCBEAM}. A continuous $\mathrm{\bar{p}}$~beam of 150~eV energy, lasting for about 20~s, was extracted and transported through the positron-antiproton mixing trap to the annihilation target, as shown in Fig.~\ref{fig:whole_setup}. The magnetic field of the double cusp, used for production of spin-polarized beam for the antihydrogen experiment, was set to the nominal value~\cite{double_cusp_Nagata_2014, cusp_Hbar_kuroda}.

The antiprotons were annihilated on three different targets: 2~${\upmu \mathrm{m}}$ thick diamond-like-carbon (DLC) and molybdenum foils, and a gold foil of 1~${\upmu \mathrm{m}}$ thickness, biased at 1~kV. The stopping range of 1.15~keV antiprotons, according to ref.~\cite{pbar_keV_stopping_power}, is $\sim$46~nm for carbon and $\sim$57~nm for gold. The beam was centered and focused on the target with an Einzel lens (see Fig.~\ref{fig:whole_setup}), where both the entrance and exit electrodes have additional diagonal cuts. Consisting of five electrodes, this lens allows for two-dimensional steering perpendicularly to the beam axis.

\begin{figure}
  \begin{minipage}{\linewidth}
    \centering
    \includegraphics[width=0.8\linewidth]{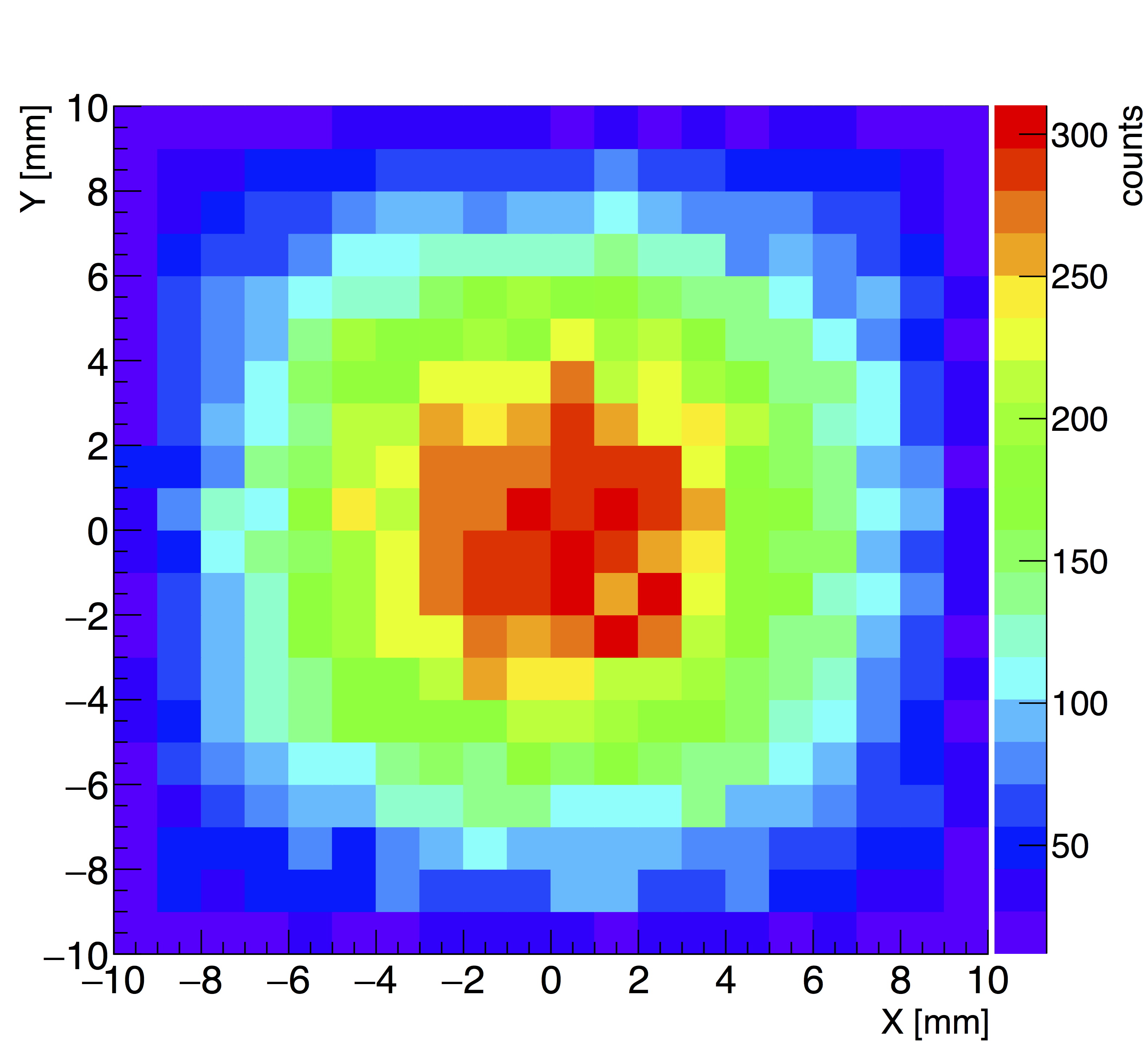}
        \subcaption{}
  \end{minipage}%
  
  \begin{minipage}{\linewidth}
    \centering
    \includegraphics[width=0.8\linewidth]{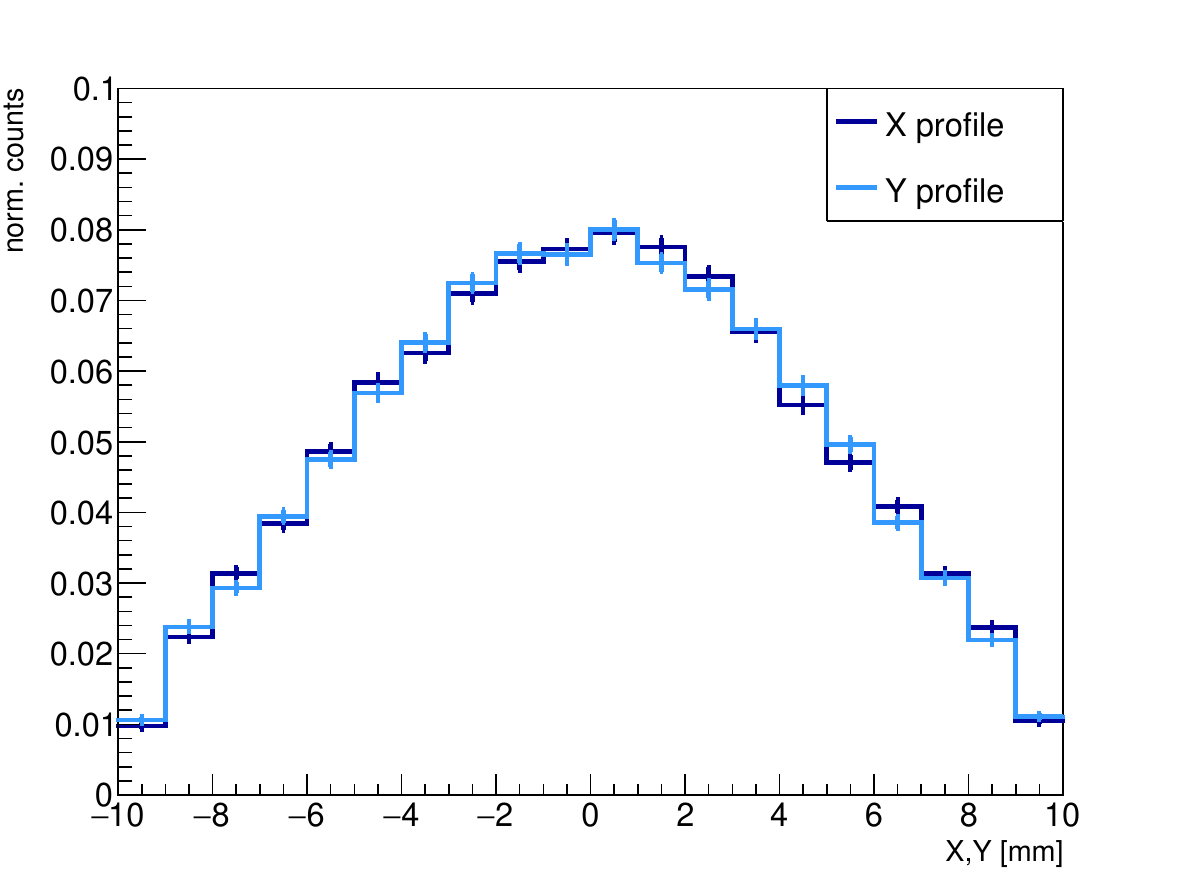}
        \subcaption{}
  \end{minipage}
  \caption{a) x-y profile of the antiproton beam hitting the target foil, obtained through measurements and Monte Carlo simulations combined (see the text for details). A total of 50,000 events were simulated. b) Projections of the beam profile in x (horizontal) and y (vertical).} 
  \label{fig:XY_pbar_beam_trig_time}
\end{figure}

The on-target beam size was determined by first measuring the antiproton beam on the bismuth germanium oxide (BGO) crystal detector used for the antihydrogen experiment \cite{BGO_Nagata_2016}, and by using the results of $\sim$30~mm FWHM in x and y direction~\cite{Widmann2300138} as an input to simulations. As the foils then replaced the BGO for the current measurements, the beam size on the target foils was estimated from the measured size of the beam, adding, in the Geant4 simulations, the 1~kV electric potential applied on the foil for further focusing. The outcome of this combined work is shown in Fig.~\ref{fig:XY_pbar_beam_trig_time}a and \ref{fig:XY_pbar_beam_trig_time}b, where the x-y beam profile and its horizontal (x) and vertical (y) projections are shown, respectively. A beam size of 10.5~mm in x and y (FWHM) was obtained, with 92.4$\%$ of the extracted antiprotons annihilating on the 2$\times$2~$\mathrm{cm^{2}}$ target, and the remainder on the mechanical support. These results were also confirmed with SIMION~\cite{simion} simulations, starting from the same initial conditions and including the voltage bias on the foil.

\begin{figure*}[htbp]
    \centering
    \includegraphics[width=0.9\textwidth]{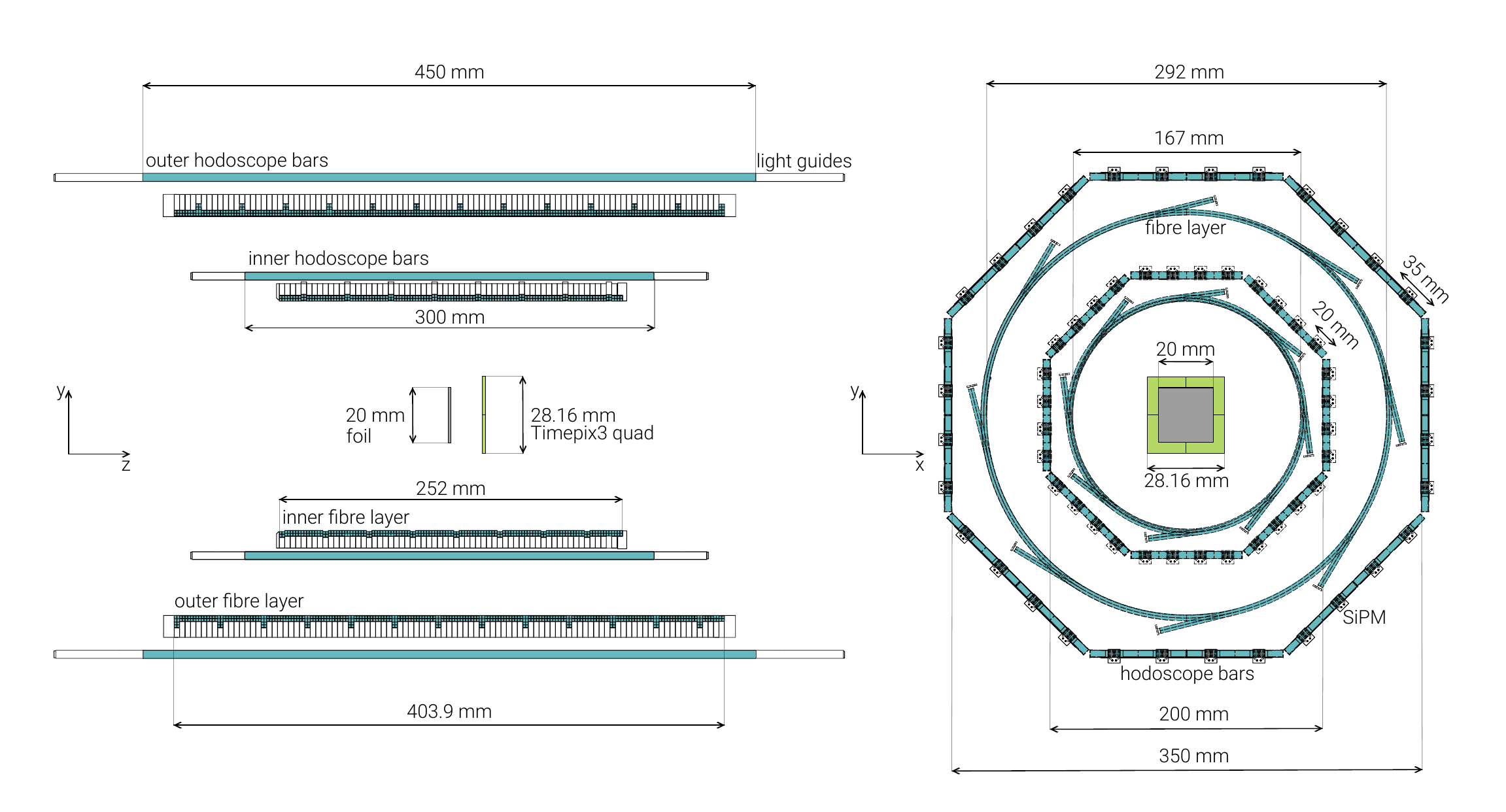}
    \caption{Schematic overview of the hodoscope, the annihilation foil and the Timepix3 detector with the corresponding dimensions.}
    \label{fig:tpx_setup}
\end{figure*}

\subsection{\label{subsec:detection}Detection and identification of charged pions and heavy fragments}
The charged annihilation products were simultaneously detected by two detectors, a cylindrical hodoscope surrounding the target and a pixel detector placed 1 cm behind it (Timepix3 quad), as shown in Fig.~\ref{fig:tpx_setup}. Pions and heavier fragments (p, d, t, He, Li, etc.) in the forward direction were impinging on the 2$\times$2 array of Timepix3, with a total area of 2.8$\times$2.8~$\mathrm{cm^{2}}$. The hybrid pixel detector (HPD) readout chip, containing 256$\times$256 pixel channels of 55$\times$55~${{\upmu \mathrm{m}}^2}$ each, provides a simultaneous information of the time-of-arrival (ToA) and the time-over-threshold (ToT) for each pixel, with a nominal time resolution of 1.56~ns. The choice of the sensor depends on the particular application, and, as silicon is well suited for detection of charged particles, the quad was coupled to a 500~${\upmu \mathrm{m}}$ thick silicon sensor. The detector is able to detect mixed radiation fields, composed of various types of particles, and has a long history of a variety of applications~\cite{tpx3_rad_imaging,timepix3_heavy_ions_spectro,timepix3_3Dtracks, Bergmann2019,Gohl2016}. Its wide dynamic spectral range allows for detection of all charged particles depositing energy above the threshold. In this experiment, the detection threshold was set to $\sim$$\mathrm{1,000~e^{-}}$ of collected charge per pixel, corresponding to a minimum required energy of $\sim$3.6 keV for X rays and low-energy gamma rays, tens of keV for electrons, hundreds of keV for protons and $\sim$MeV for heavy ions~\cite{GRANJA201860}. The higher threshold for heavier particles originates from the aluminum electrode, approximately 500~nm in thickness, that is deposited on top of the silicon sensor for biasing purposes. For details about the ASIC and the readout system see ref.~\cite{timepix3, medipix3,timepix3_SPIDR}.

Omnidirectional charged pions were detected with the hodoscope encircling the foil (Fig.~\ref{fig:tpx_setup}), consisting of two layers of scintillating plastic bars (material EJ-200) parallel to the beam direction, and two layers of scintillating fibres (material Saint Gobain BCF-12) perpendicular to the beam and encircled by each bar layer. The widths of the bars in the inner and outer layer are 20~mm and 35~mm, respectively, with a thickness of 5~mm. Both bar layers have an octagonal x-y cross section and are composed of eight panels, each composed of four scintillating bars. Every bar is read out on both ends by silicon photomultipliers (SiPMs, KETEK 3350TS) of 3$\times$3~$\mathrm{{mm}^2}$ active surface. The bar hodoscope is capable of measuring the time-of-flight (ToF) for particles crossing both layers~\cite{asacusa_hodo}. The fibres, forming two layers of concentric cylinders are also read out with SiPMs (KETEK PM3350-EB), bonded with optical cement on one end of each fibre bundle. The fibre part of the hodoscope provides an increased position resolution in z-direction. A detailed description is given in ref.~\cite{kolbinger_thesis}.

\begin{figure*}[h]
  \begin{minipage}{0.5\textwidth}
    \centering
    \includegraphics[width=\linewidth]{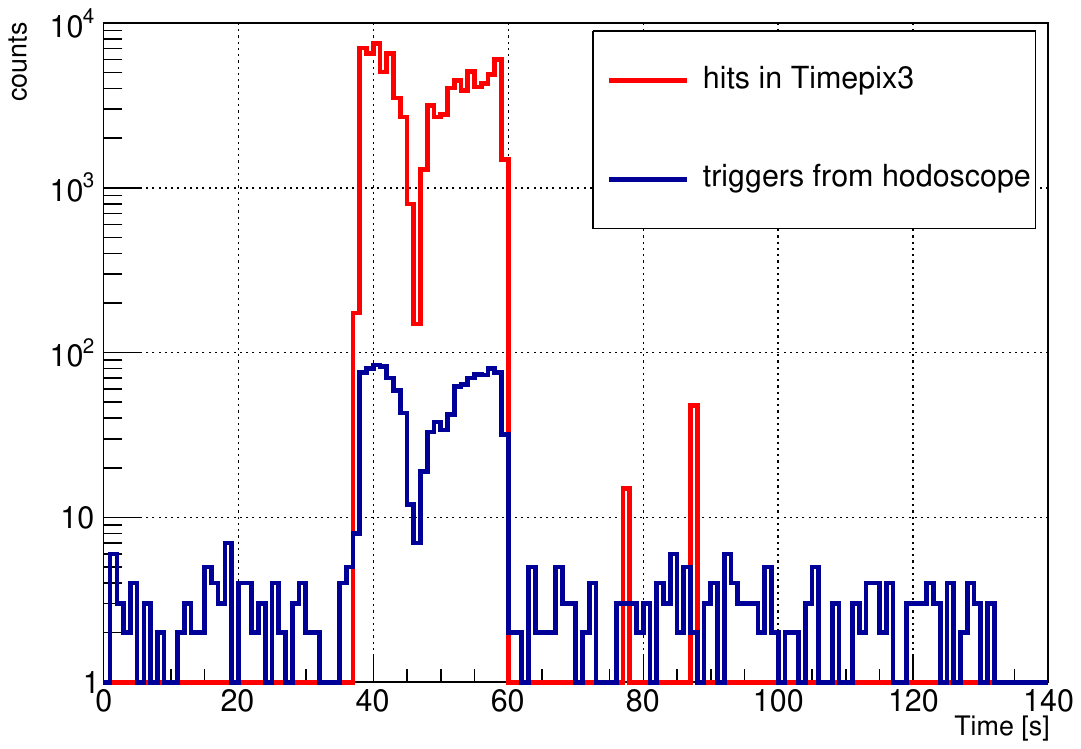}
    \subcaption{}
  \end{minipage}%
  \begin{minipage}{0.5\textwidth}
    \centering
    \includegraphics[width=\linewidth]{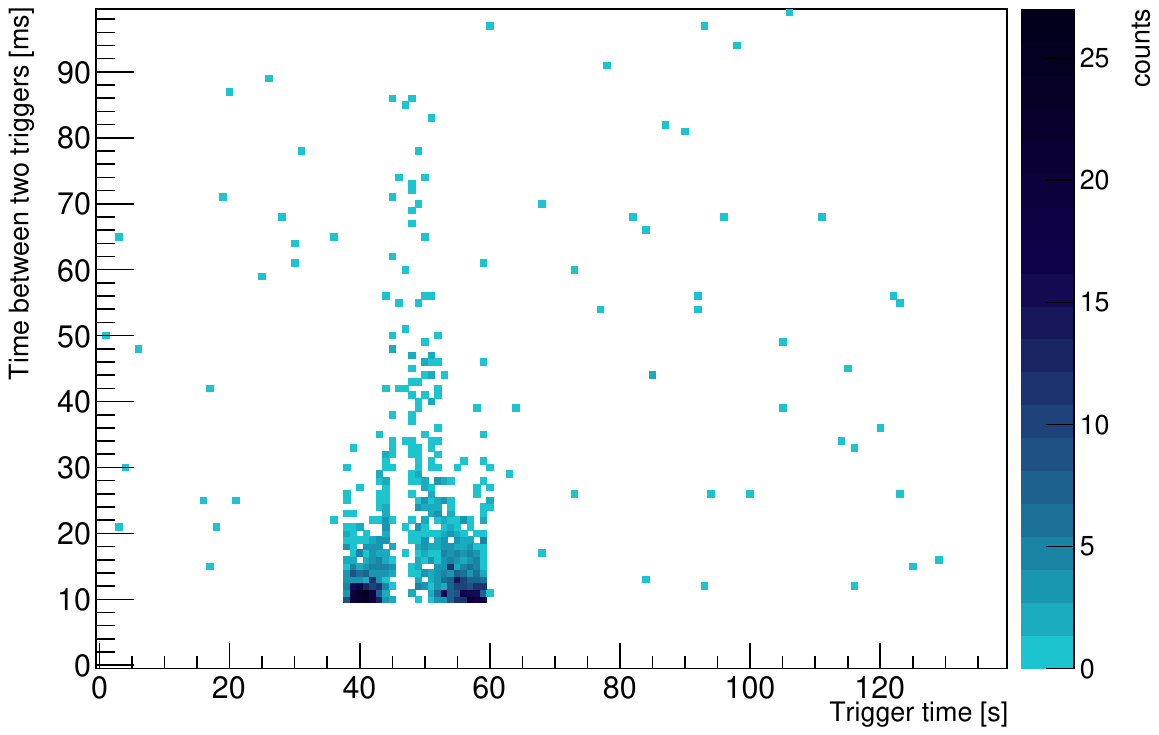}
    \subcaption{} 
  \end{minipage}
  \caption{a) A sample time distribution from one extraction cycle showing the triggers issued by the hodoscope and the number of hits (individual pixels) in the Timepix3. The $\sim$20~s long slow extraction of antiprotons is clearly visible (see text for more details). b) The time between two consecutive triggers in ms plotted against the trigger time for the same extraction cycle.}
  \label{fig:triggers_hits}
\end{figure*}

The two detectors were time synchronised, and the data acquisition coincided with the start of the slow extraction cycle. Every time a hit was detected in both inner and outer bar layers of the hodoscope, a trigger was issued and this event was time-stamped in the Timepix3 data stream, where every hit in this detector was also recorded. The time distribution of these events (triggers) for one extraction cycle is shown in Fig.~\ref{fig:triggers_hits}a, revealing the 20~s duration of the antiproton beam extracted from MUSASHI within the 140~s long cycle. The double peak structure is related to the DC extraction, and possible sources are the imperfect cooling of the antiprotons, or a mismatch to the space charge potential when the well depth was slowly decreased during the extraction. About 1,000 $\mathrm{\bar{p}}$~annihilation events per slow extraction cycle were detected, resulting in a total of $\sim$100,000 annihilation events for each target foil. The triggers outside the slow extraction window are produced by cosmic rays or, sporadically, from secondaries of other origin in the AD experimental hall crossing the hodoscope. The two small peaks in the Timepix3 distribution are from two cosmic particles that traversed both the hodoscope and the Timepix3 detectors. 
The data analysis was performed only within the slow extraction time window. 

The absence of degraders to moderate the antiprotons prevents pion contamination from upstream beam annihilations, observed in previous measurements~\cite{mimotera, emulsion1, timepix3_Helga_paper}, restricting the background contamination to essentially cosmic rays.

The hits in the Timepix3 were assigned to a particular event (trigger) according to their ToA. The interval between two successive triggers ranged from 10~ms to a few 10~ms, as illustrated in Fig.~\ref{fig:triggers_hits}b. The observed cutoff at 10~ms stems from the hodoscope data acquisition limit of 100~Hz. 
The nominal cosmic background was measured to be $\sim$1.3~Hz. When combined with the duration of one extraction (20 s) it gives a total of 26 cosmics events, which is only 2.6\% of the average number of annihilation events recorded per run. For the Timepix3 the cosmic ray background is fully negligible.

\subsubsection{Minimum ionizing particles in the hodoscope}
Minimum ionizing particles (MIPs) were detected with the hodoscope, covering $\sim$80\% of the full solid angle. The majority of these particles originate from charged pions generated in the annihilation process. However, a notable fraction of the identified tracks are generated by other particles such as electrons, positrons, and energetic protons (see Sec.~\ref{sec:MC_sims}). The $e^{+}e^{-}$ pairs arise from the conversion of the two 67~MeV $\gamma$-rays from the $\pi^{0}$ decay, or from Dalitz pair decay $\pi^{0} \rightarrow \gamma e^{+} e^{-}$. Some of the high-energy protons, with hundreds of MeV of energy originating from the final state interactions can also cross all four layers of the hodoscope and produce tracks.

A three--dimensional (3D) tracking algorithm was employed to ascertain the number of MIPs for individual annihilation events. The algorithm groups corresponding hits in the fiber and bar detector into subsets and identifies track candidates by combining these hits. Subsequently, a linear fit is applied to all candidates within the track collection, and those candidates that best conform to a straight line are chosen. Further procedural intricacies are outlined in detail in ref.~\cite{kolbinger_thesis}. Additionally, the determined tracks are used to reconstruct the annihilation vertex. The resolution $\sigma_{z}$ in the z-direction for the vertex reconstruction is $\sim$50~mm.

\subsubsection{Charged pions and heavier fragments in Timepix3} \label{subsec:prongs_in_tpx3}
\begin{figure}[htp]
  \begin{minipage}{\linewidth}
    \centering
    \includegraphics[width=\linewidth]{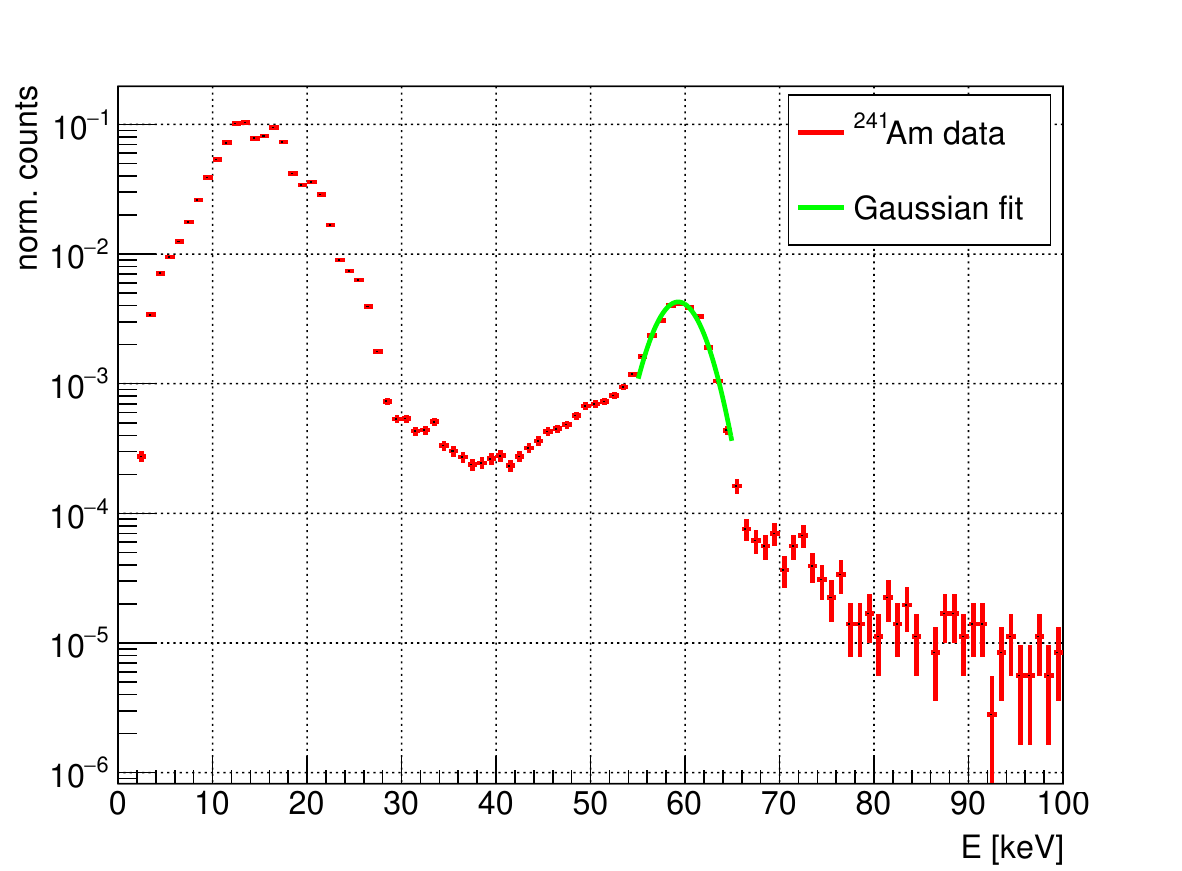}
    \subcaption{}
  \end{minipage}

  \begin{minipage}{\linewidth}
    \centering
    \includegraphics[width=\linewidth]{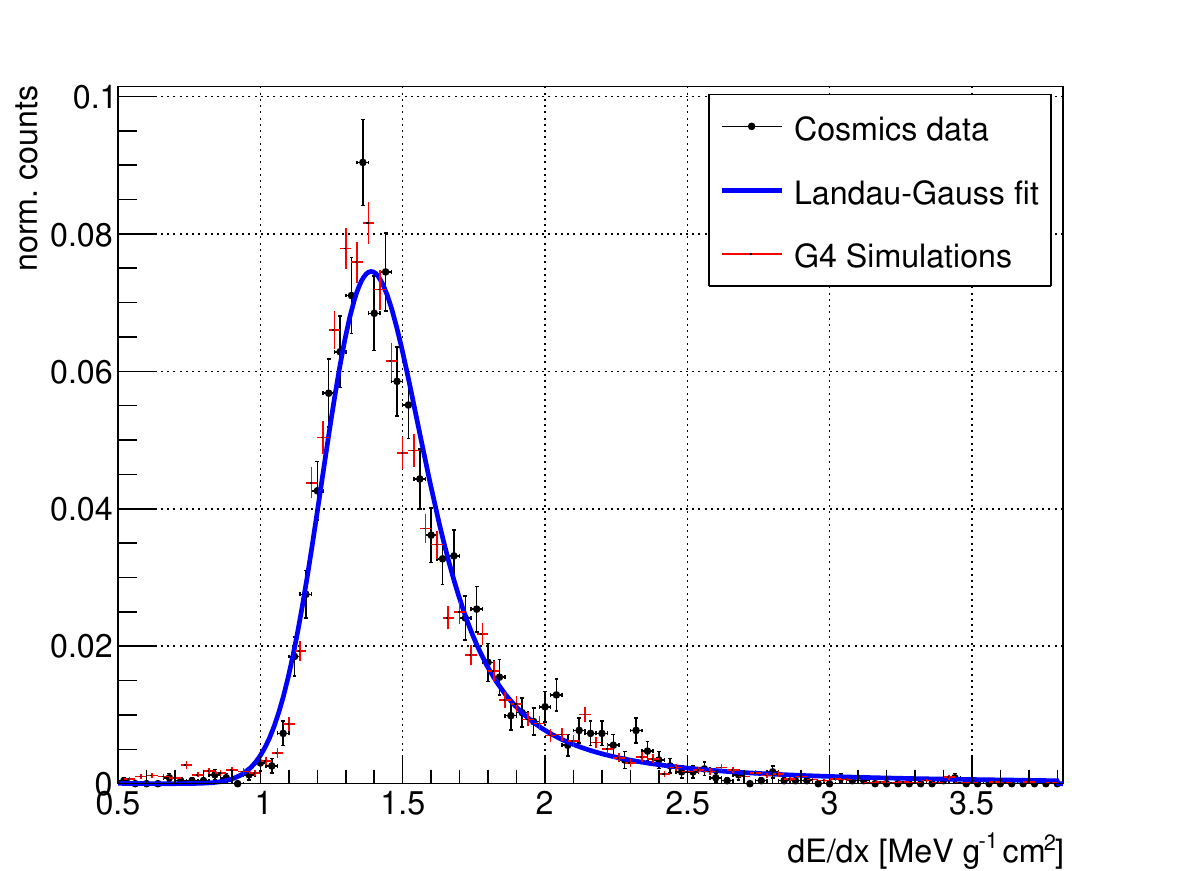}
    \subcaption{}
  \end{minipage}
   \caption{a) Energy spectra of $\gamma$-rays from $\mathrm{^{241}Am}$ measured with the Timepix3 quad. The characteristic 59.5~keV peak is shown together with a Gaussian fit. b) $dE/dx$ distribution for cosmic rays measured with the same detector and modeled with a Landau-Gaussian fit, compared to Geant4 simulations.}
  \label{fig:Am241_energy}
\end{figure}

In the pixel detector, charged particles were detected in $\sim$25\% of the full solid angle. Based on the ToT proportionality to the amount of charge collected in each electrode, a per pixel energy calibration was carried out using test pulses. The deposited energy was extracted from the known value of the test pulse capacitor, assuming an energy of 3.6~eV for the creation of one electron-hole pair in silicon. The calibration results were validated with data from a $\mathrm{^{241}Am}$ radioactive source and cosmic rays. 
In Fig.~\ref{fig:Am241_energy}a, the lower end of the $\mathrm{^{241}Am}$ spectrum, converted to energy units, is presented with a Gaussian fit to the characteristic 59.5~keV peak, determined to be 59.3$\pm$2.6~keV. 

The energy deposit and morphology of the pixelated track that a charged particle generates in the Timepix3 (also referred to as pixel cluster) are influenced by factors such as the particle type, its kinetic energy, and the incidence angle. The graph in Fig.~\ref{fig:Am241_energy}b illustrates the $dE/dx$ distribution for cosmic rays, both in measurements and simulations, exclusively for clusters with high ($>$0.8) eccentricity, a measure of the linearity of a particle track. These elongated tracks correspond to cosmic rays that cross the Timepix3 at large angles with respect to its normal. The distribution is modeled with a Landau-Gaussian fit and compared to Geant4 simulations. The most probable value of $\sim$1.4~$\mathrm{{MeVg^{-1}cm^{2}}}$ agrees with the typical value characteristic for MIPs in silicon~\cite{Bergmann2020}.

Due to the geometry of the set-up (see Fig.~\ref{fig:tpx_setup}), the annihilation prongs hit the pixel detector only at acute angles with respect to its normal. The maximum angle is $\sim$55$\degree$ for prongs emerging from the foil's center, and up to $\sim$67$\degree$ for prongs originating from annihilations close to the edges of the target. Most of the particles impinge on the Timepix3 at angles smaller than these, which results in the formation of mostly blob-like clusters and relatively short tracks, with a maximum length of less than 1.5~mm. Combined with the variety of particles produced in the annihilation and their diverse ranges of energies and incident angles, this allowed only a broad classification of heavily ionizing particles (HIPs), as opposed to MIPs. In this work, protons and all heavier nuclear fragments are classified as HIPs, regardless of their energy. Ongoing efforts from a number of users of Timepix ASICs are aimed at developing tools for more precise particle identification in mixed radiation fields~\cite{GRANJA201860}. 

\begin{figure}[htp]
  \begin{minipage}{\linewidth}
    \centering
    \includegraphics[width=\linewidth]{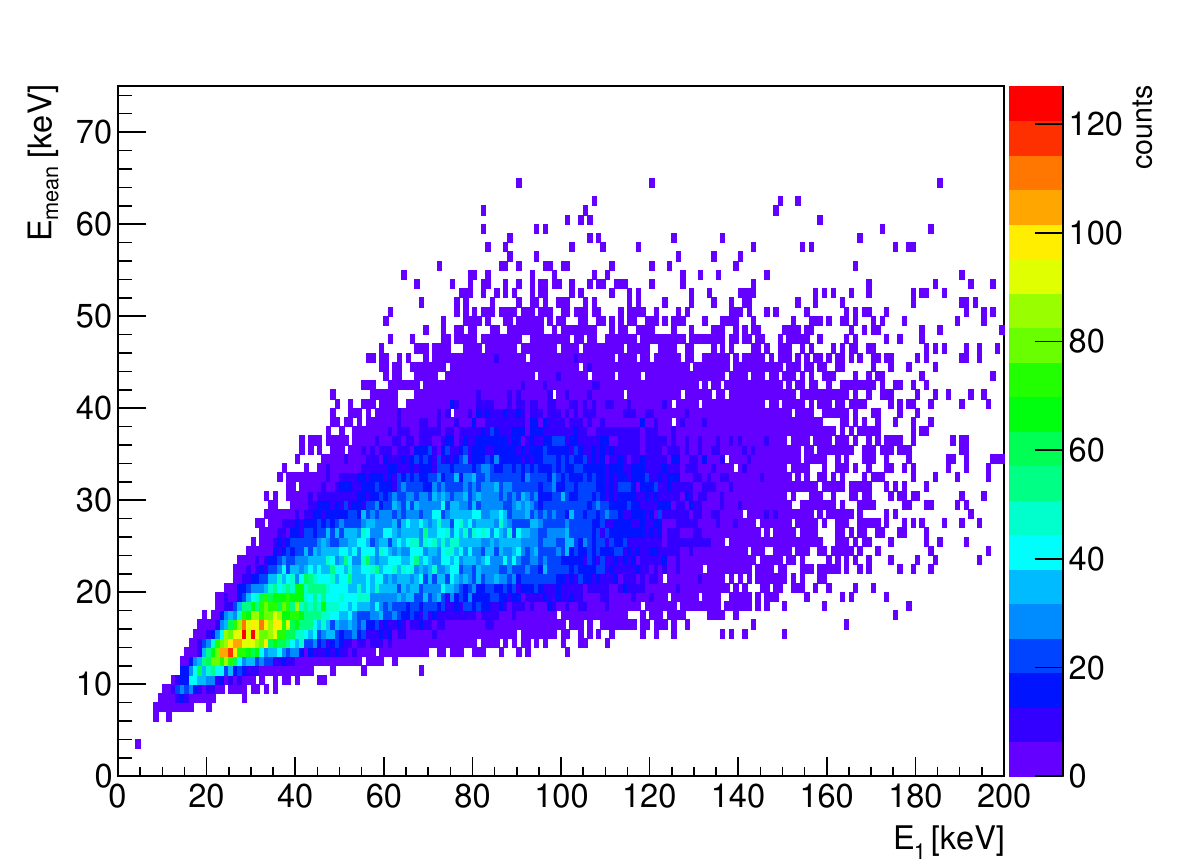}
    \subcaption{}
  \end{minipage}

  \begin{minipage}{\linewidth}
    \centering
    \includegraphics[width=\linewidth]{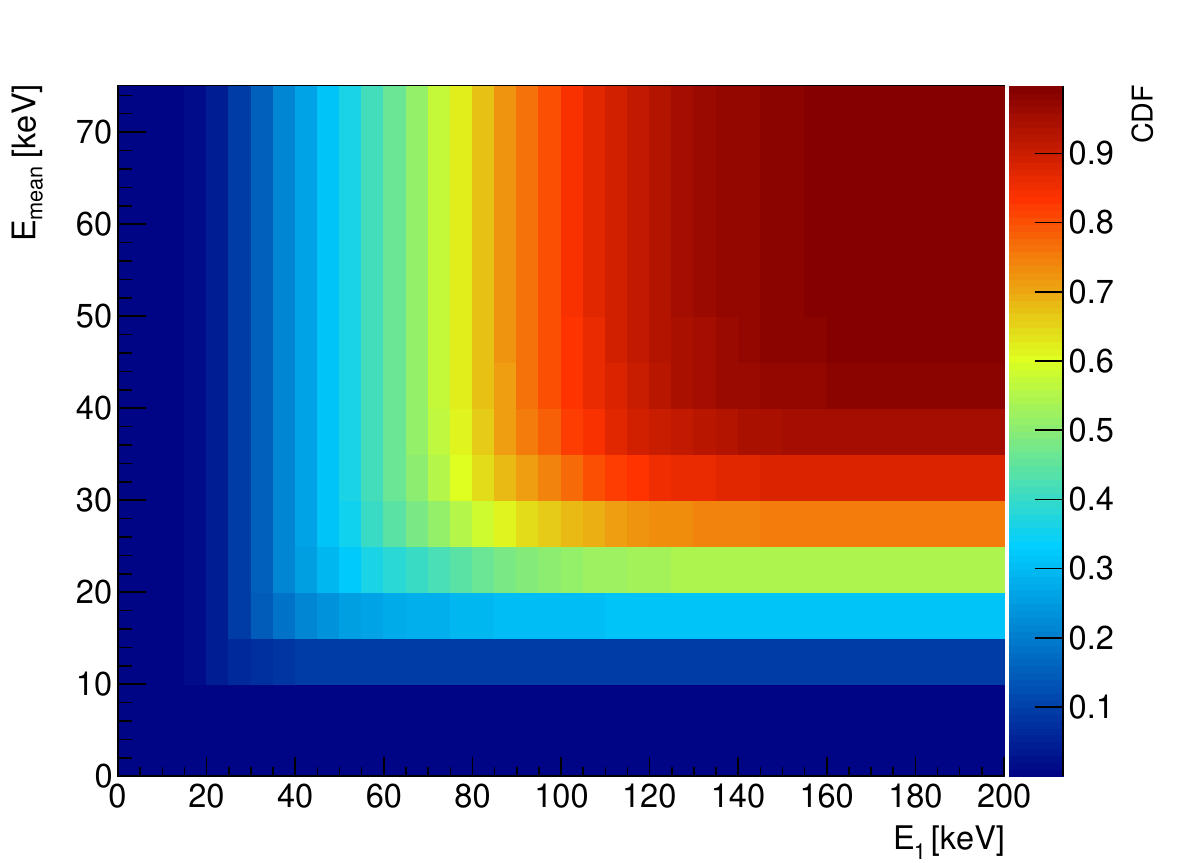}
    \subcaption{}
  \end{minipage}
   \caption{a): Mean pixel energy ($E_{\mathrm{mean}}$) versus highest pixel energy ($E_{\mathrm{1}}$) for clusters produced by cosmic rays in Timepix3. b) Cumulative distribution function (CDF) for the above 2D distribution, used to extract cuts on $E_{\mathrm{mean}}$ and $E_{\mathrm{1}}$, to exclude MIPs from the antiproton annihilation data.}
  \label{fig:cosmics_cuts}
\end{figure}

\begin{figure}[htp]
  \begin{minipage}{\linewidth}
    \centering
    \includegraphics[width=\linewidth]{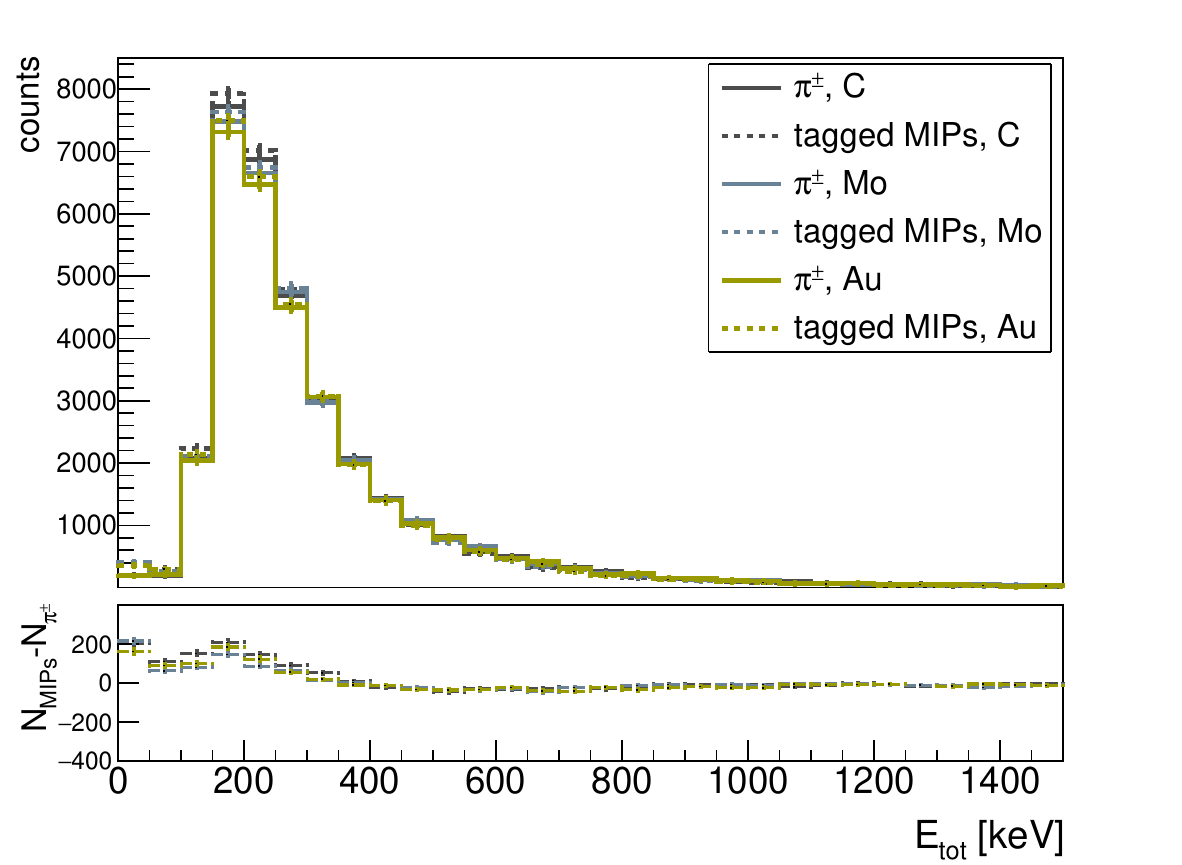}
    \subcaption{}
  \end{minipage}

  \begin{minipage}{\linewidth}
    \centering
    \includegraphics[width=\linewidth]{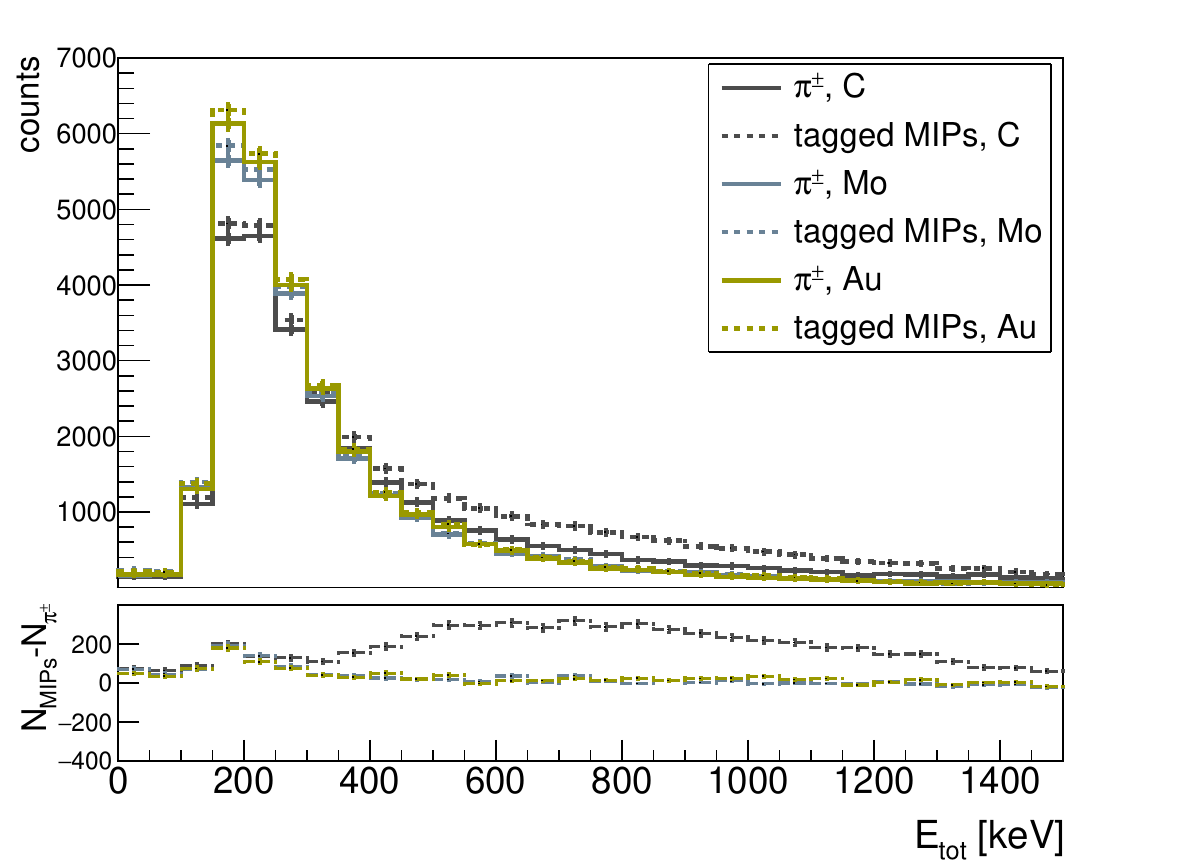}
    \subcaption{}
  \end{minipage}
   \caption{Deposited energy in Timepix3 from MIPs created in $\mathrm{\bar{p}}$ annihilations with C, Mo and Au nuclei, simulated with a) FTFP and b) CHIPS models in Geant4. The distributions of the charged pions are shown with solids lines, whereas the dashed lines are for MIPs tagged by applying energy cuts from simulated cosmic rays (as described in the text). The bottom panels show the difference between $\pi^{\pm}$ and MIPs for each foil.} 
  \label{fig:Etot_MIPs_tag}
\end{figure}

\begin{figure}[htp]
  \begin{minipage}{\linewidth}
    \centering
    \includegraphics[width=\textwidth]{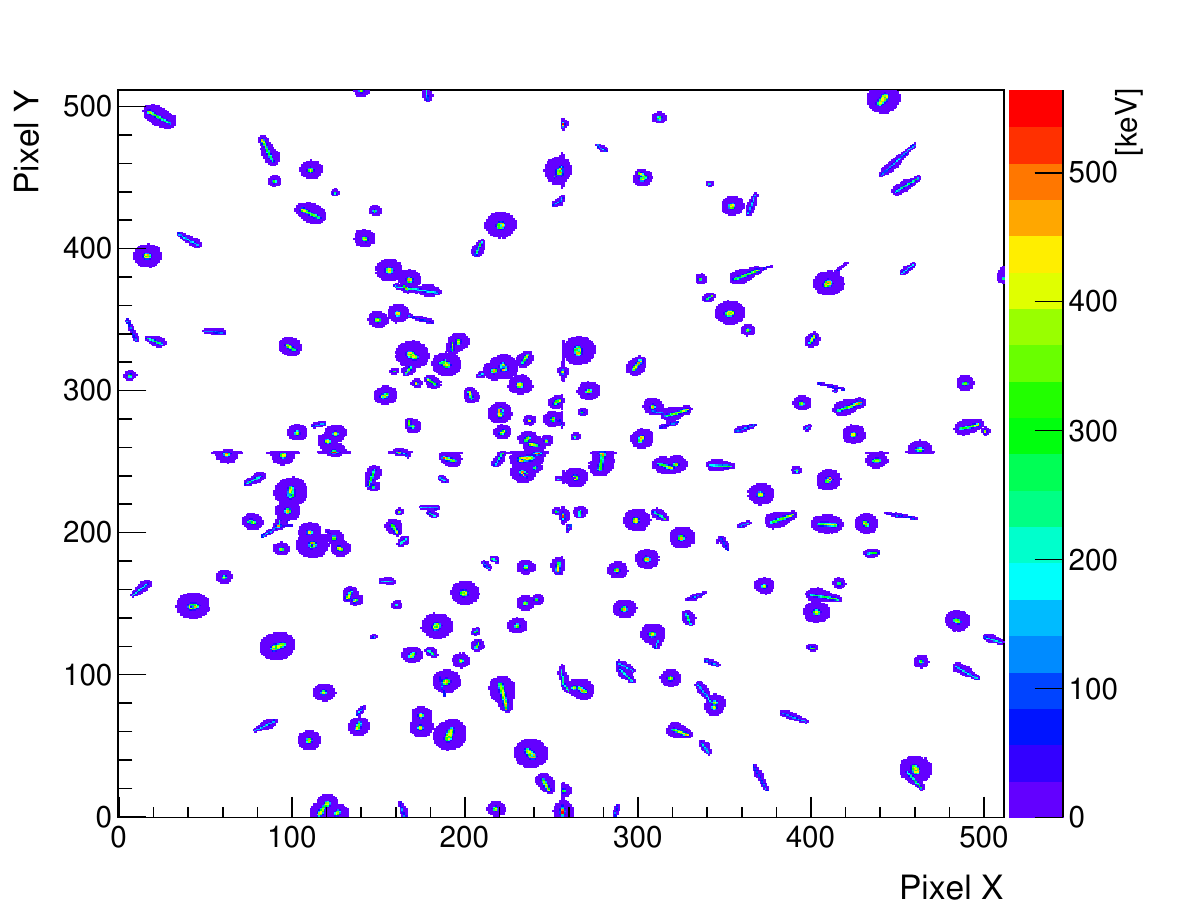}
    \subcaption{}
  \end{minipage}

  \begin{minipage}{\linewidth}
    \centering
    \includegraphics[width=\textwidth]{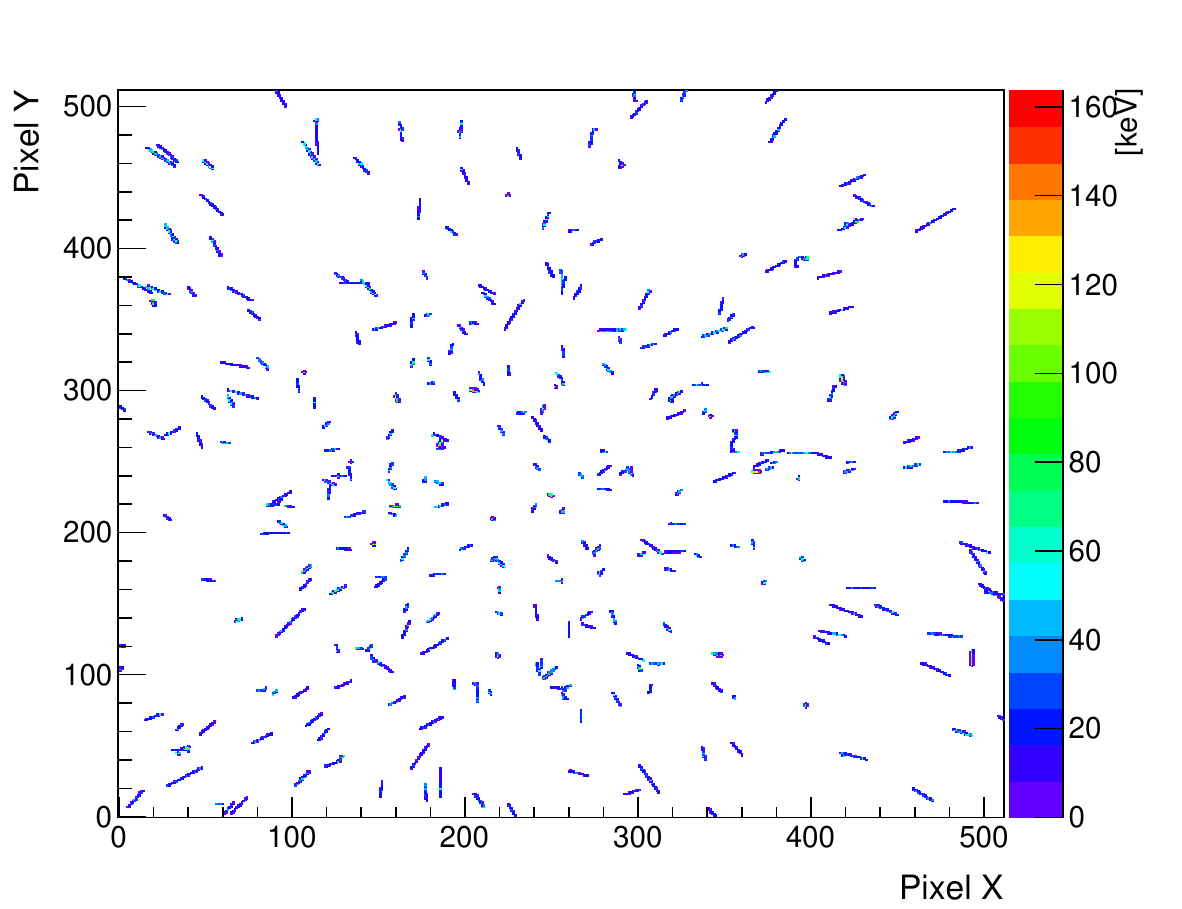}
    \subcaption{}
  \end{minipage}
    \caption{Hitmaps of prongs from 280 antiproton annihilations in carbon, after their categorisation in a) HIPs and b) MIPs. The colour scale shows the deposited energy in keV/pixel.}  
  \label{fig:hitmaps_tpx3}
\end{figure}

In this study, the differentiation of MIPs from HIPs in the Timepix3 was achieved by studying the cluster morphology and energy deposits from a cosmics data sample. Since the distribution of the energy deposit from a cosmic, or a MIP in silicon partly overlaps with that of protons with energies of $\sim$100~MeV, other criteria, more specific to the cosmics particles were required to distinguish MIPs in the annihilation data. Hence, as a first step, a double cut was derived from the analysis of cosmic data on the highest pixel energy ($E_1$) within the cluster and the mean pixel energy ($E_{\mathrm{mean}}$). The energy thresholds were extracted from the $E_{\mathrm{mean}}$ versus $E_1$ distribution depicted in Fig.~\ref{fig:cosmics_cuts}a, by calculating its cumulative distribution function (CDF), shown in Fig.~\ref{fig:cosmics_cuts}b. The thresholds were set to account for $>$99$\%$ of the cosmic rays data sample, leading to values of 55~keV and 164~keV for $E_{\mathrm{mean}}$ and $E_1$ respectively. The remaining clusters underwent further evaluation based on their morphology and were categorized as small blobs, large blobs, or heavy tracks before being identified as HIPs (see Fig.~\ref{fig:hitmaps_tpx3}a).
The efficiency of removing MIPs from the annihilation clusters using only energy cuts from cosmics was estimated, from simulations, to be $\sim$97$\%$. However, it was also revealed that some high-energy protons can be mistakenly identified as MIPs after applying these cuts, as shown in Fig.~\ref{fig:Etot_MIPs_tag}. The comparison plots show the deposited energy by charged pions (solid line) and by MIPs tagged after applying the energy cuts obtained from simulated cosmic rays (dashed lines) for two Geant4 models ((a) FTF and (b) CHIPS), for all three targets. Additionally, the difference between the two energy depositions is presented at the bottom. The excess counts are attributed to the aforementioned protons, the quantity of which depends on the specific simulation model. The percentage of these protons incorrectly tagged as MIPs relative to the total number of detected HIPs varies between 2\% for Mo in CHIPS to 15\% for C and Au in FLUKA. 

In data, this issue was addressed by introducing an additional cut on the number of halo pixels (explained in Sec.~\ref{sec:MC_sims}), extracted from the cosmic data in a similar way as the energy cuts. This way, blob-like and heavy tracks produced by protons, were distinguished among the clusters initially tagged as MIPs with the energy cuts. The efficiency of the cut on the halo pixels could not be assessed in simulations, since the halo is an ASIC-specific effect present only in the measured data. However, the extracted HIPs after the implementation of this cut in data amount to 10\% of the total number of HIPs in C and Mo, and to 12\% in gold. These values are well within the limits of the model's predictions for protons incorrectly tagged as MIPs. 
In Fig.~\ref{fig:hitmaps_tpx3}a, the hitmap shows clusters from 280 annihilation events that were classified as HIPs, while Fig.~\ref{fig:hitmaps_tpx3}b shows those discarded as MIPs.

\section{\label{sec:MC_sims}Monte Carlo simulations}

The Monte Carlo simulations were performed using three of the physics lists available in Geant4, as well as FLUKA (fluka 4-2.1). The $\mathrm{QGSP\_BERT\_CHIPS}$ model was simulated in Geant4.9.6.p04, $\mathrm{FTFP\_BERT\_EMY}$ in Geant4.10.05.p01, and the most recent one, $\mathrm{FTFP\_INCLXX\_EMZ}$ in Geant4.11.2.1. The four models adopt different approaches when describing antiproton-nucleus annihilation at rest. The presented comparisons comprise the aforementioned physics lists in their integral form, but in future this work could be used to compare the relative importance of different effects.

In CHIPS, any excited hadronic system, such as an incident particle and target nucleons or nuclear matter, is considered to be a quasmon - a bubble containing massless quarks (quark-parton plasma). The fragmentation of a quasmon into hadrons occurs through a quark fusion mechanism~\cite{CHIPS_pbarp}. This is how the mesons are produced when the antiproton annihilates on a peripheral nucleon. The final state interactions that follow between these mesons and the residual nucleus are treated in a similar way, by generating quasmons whenever some of these mesons are absorbed by the nucleus. The CHIPS model, as opposed to INC, does not develop cascades, and the final state particles are created through hadronization of the quasmons inside the same nucleus~\cite{Kossov2005}.

The underlying idea of Fritiof is based on the hadronic string approach which is used in Monte Carlo event generators like Pythia~\cite{pythia}. Here it is combined with the Glauber theory for nucleus-nucleus interactions, which is used to calculate cross sections for processes that involve the nucleus, such as the antiproton-nucleus annihilation. The initially created string between the antiproton and the target nucleus breaks up into several clusters, which subsequently decay into final-state particles. Even though it can simulate the production of intermediate resonances during the annihilation process, which adds to the understanding of the underlying nuclear structure, the model does not account for the FSI between the primarily produced mesons and the residual nucleus, which is important for predicting the multiplicity distributions of the final-state particles.

In FLUKA, $\mathrm{\bar{p}}$A annihilation at rest starts with the antinucleon-nucleon process, modeled through the production and decay of two or more intermediate states whose branching ratios are adjusted to reproduce multiplicities for pions, kaons and resonances from experiments. The subsequent interaction of these particles with the remaining nucleus, as well as all nuclear effects are treated by the custom preequilibrium cascade (PEANUT) model.

The latest update in the Geant4 simulation toolkit is the recent extension of the Intranuclear Cascade de Li\`ege (INCL) model~\cite{INCL_87Cugnon, INCL_original} for $\mathrm{\bar{p}}$A annihilation at rest~\cite{demid_thesis_incl}. The frequencies and types of the primary annihilation products are determined from experimental data. The subsequent intranuclear cascade is based on binary collisions, particle decay during flight and interactions at the nuclear surface. The classical approach in INCL can result in the generation of states that are not physically possible, such as those prohibited by the Pauli principle. This is resolved by applying Pauli-blocking tests to the generated collisions.

\begin{figure*}
  \centering
  \begin{minipage}{0.45\textwidth}
    \centering
    \includegraphics[width=\textwidth]{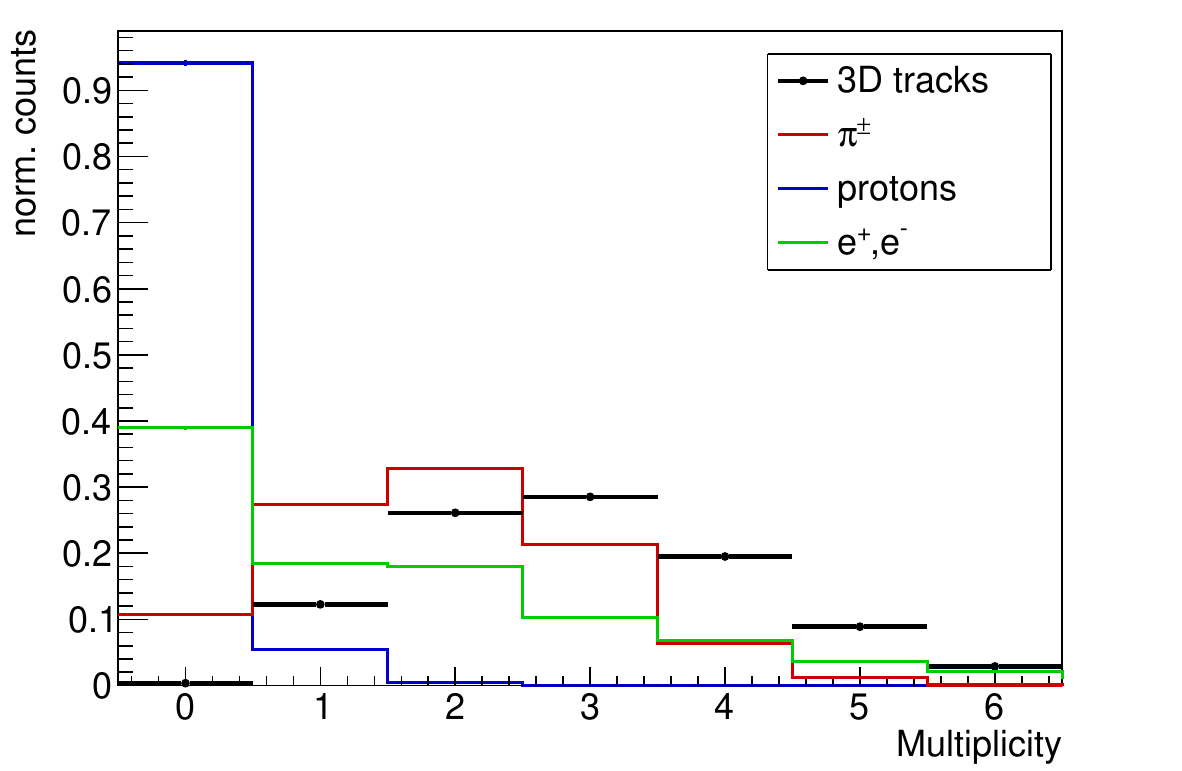 }
\subcaption{} 
  \end{minipage}\hfill
  \hfill
  \begin{minipage}{0.45\textwidth}
    \centering
    \includegraphics[width=\textwidth]{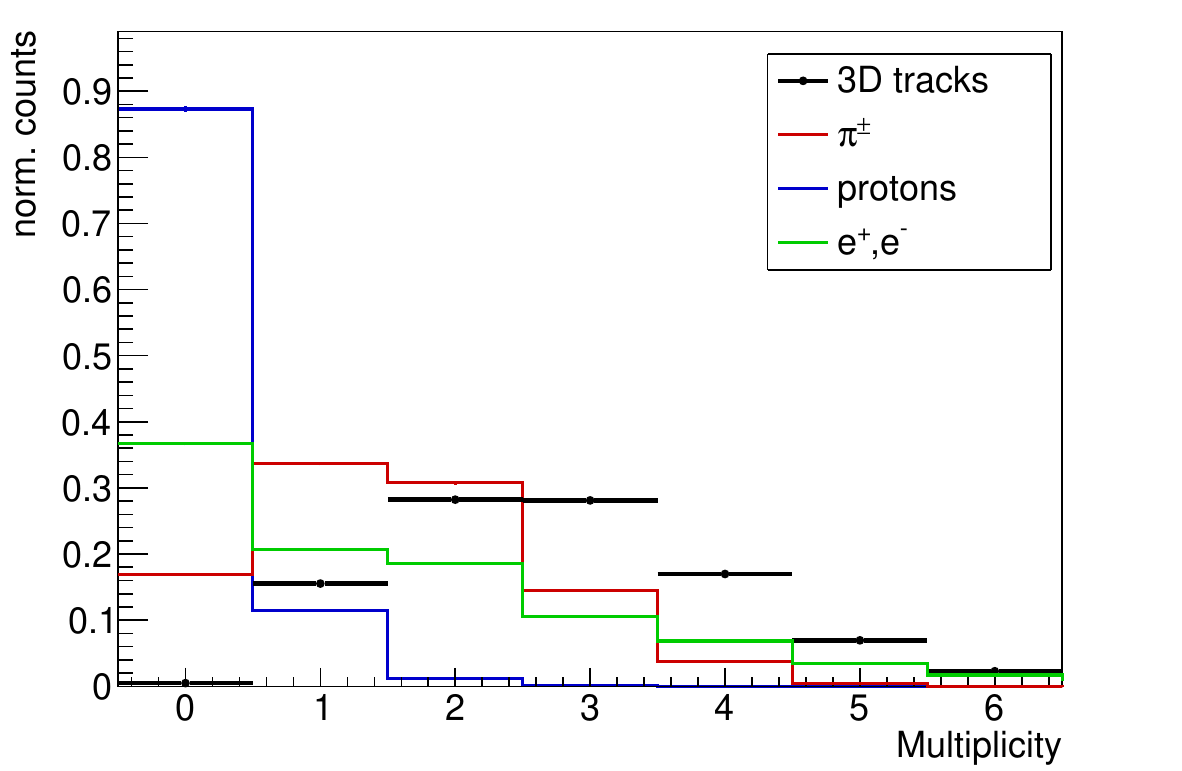}
    \subcaption{} 
  \end{minipage}\hfill

  \begin{minipage}{0.45\textwidth}
    \centering
    \includegraphics[width=\textwidth]{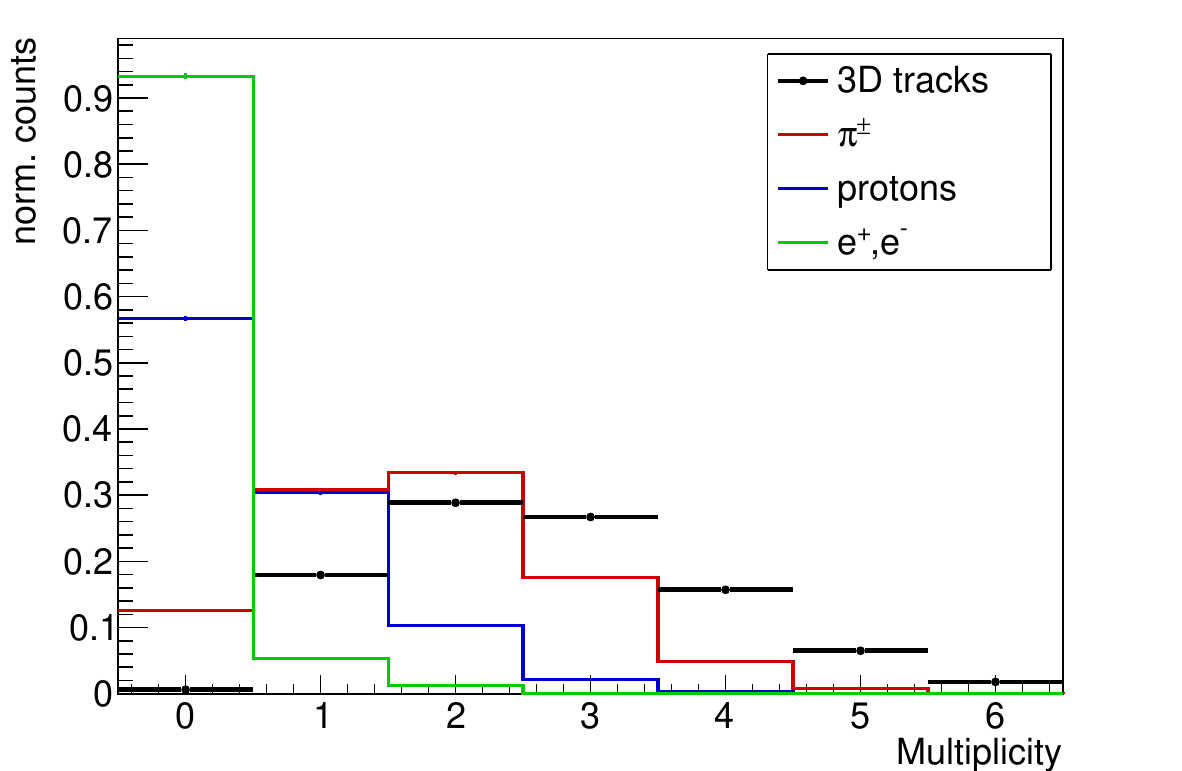}
    \subcaption{}
  \end{minipage}\hfill
  \hfill
  \begin{minipage}{0.45\textwidth}
    \centering
    \includegraphics[width=\textwidth]{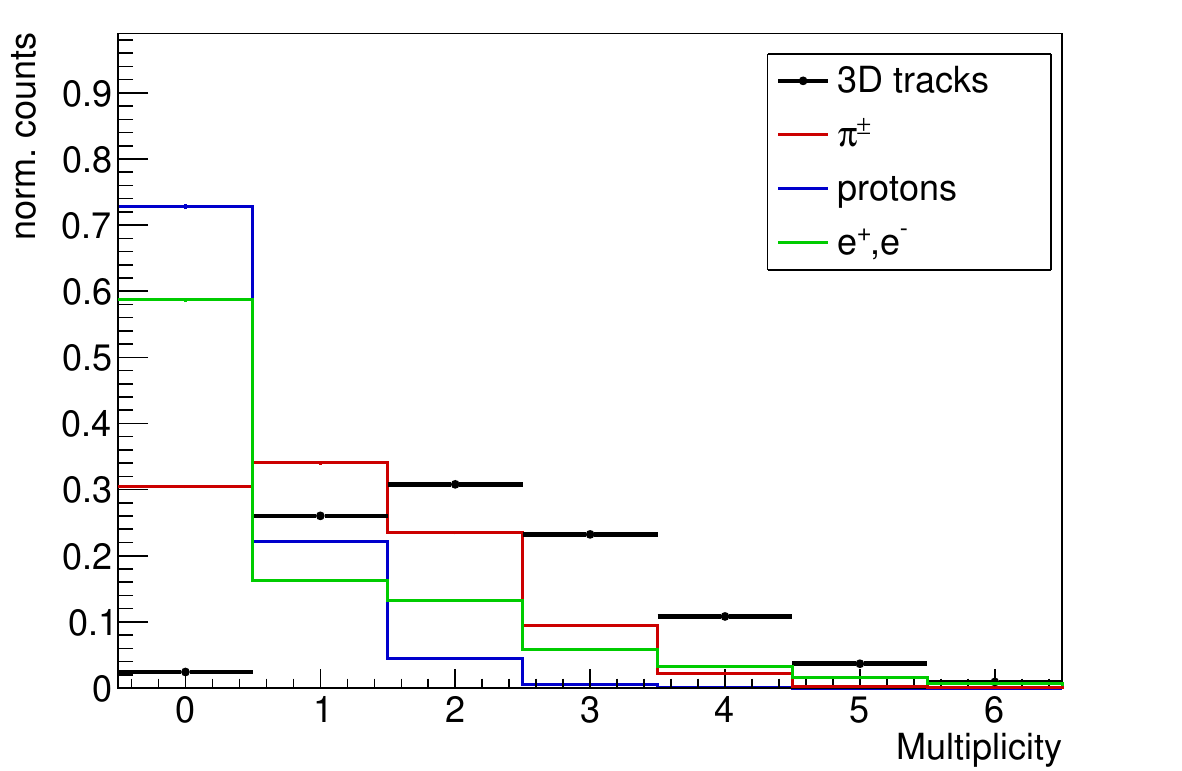}
    \subcaption{}
  \end{minipage}\hfill
   \caption{Multiplicity distributions of 3D reconstructed tracks from MIPs detected by the hodoscope in Monte Carlo simulations, compared to the number of various particles crossing the hodoscope per annihilation event, for antiproton annihilation at rest in gold with a) FTFP, b) CHIPS, c) FLUKA and d) INCL.} 

\label{fig:MIPs_distros_gold_sims}
\end{figure*}

The simulations presented here include a complete geometry of the set-up with the Timepix3 and hodoscope detectors down to their smallest components, as well as vacuum pipes, various support structures etc. To ensure a direct comparison, identical analysis tools were applied to both data and simulations. Achieving this required extracting the same type of signal from the simulations--specifically, converting the step-based energy deposits characteristic of Geant4 and FLUKA into hits in the hodoscope bars and fibers, and into pixels and clusters in the Timepix3 quad. 

For the hodoscope a straightforward threshold on the deposited energy was implemented to define a hit (0.7~MeV for the bars and of 0.55~MeV for fibres), determined from the deposited energy by simulated cosmic rays. Both Geant4 and FLUKA simulations consistently indicate that most of the tracks in the hodoscope are from charged pions. However, a significant portion is ascribed to other high--energy particles crossing the four layers of the hodoscope. In FTFP, CHIPS and INCL their contribution remains constant at $\sim$35\%, $\sim$50\% and $\sim$40\% respectively, irrespective of the target. In contrast, for FLUKA the values range between $\sim$25\% and $\sim$35\%, increasing with the mass of the nucleus. In CHIPS and FTFP models $\sim$95\% of the non-pion tracks are generated by electrons or positrons, each with tens of MeV of energy, resulting from the conversion of $\gamma$-rays emitted from the decay of $\pi^{0}$ from $\mathrm{\bar{p}}$ annihilation. In FLUKA however, the surplus of 3D reconstructed tracks is primarily attributed to high-energy protons. In INCL most of the non-pion tracks originate from electrons or positron, but proton contribution is significant with $\sim$40\%. Fig.~\ref{fig:MIPs_distros_gold_sims} shows the distribution of the number of 3D tracks per annihilation event for $\mathrm{\bar{p}}$-Au, across the four models. Additionally, the graph includes a breakdown of the particles fully traversing the hodoscope, depositing in each layer energy above the threshold that defines a hit. It is evident that the distribution of charged pions alone does not reproduce the number of 3D reconstructed tracks, underscoring the contribution of other particles crossing the detector. This becomes particularly important when discussing the multiplicity of MIPs, as it should be noted that in the annihilation process, particles other than charged pions, such as protons and electrons, can fully traverse the Hodoscope and be counted as MIPs.

\begin{figure*}
  \begin{minipage}{0.5\textwidth}
    \centering
    \includegraphics[width=\linewidth]{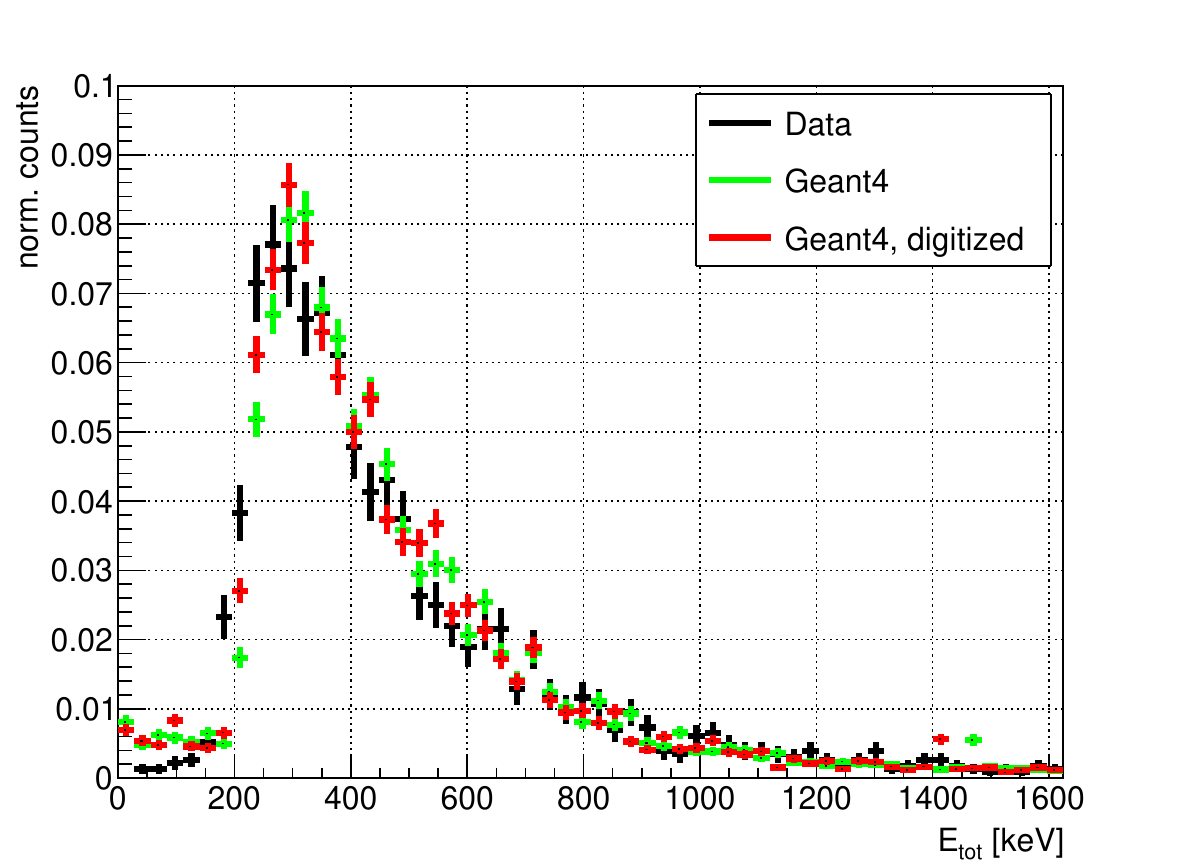}
    \subcaption{}
  \end{minipage}%
  \begin{minipage}{0.5\textwidth}
    \centering
    \includegraphics[width=\linewidth]{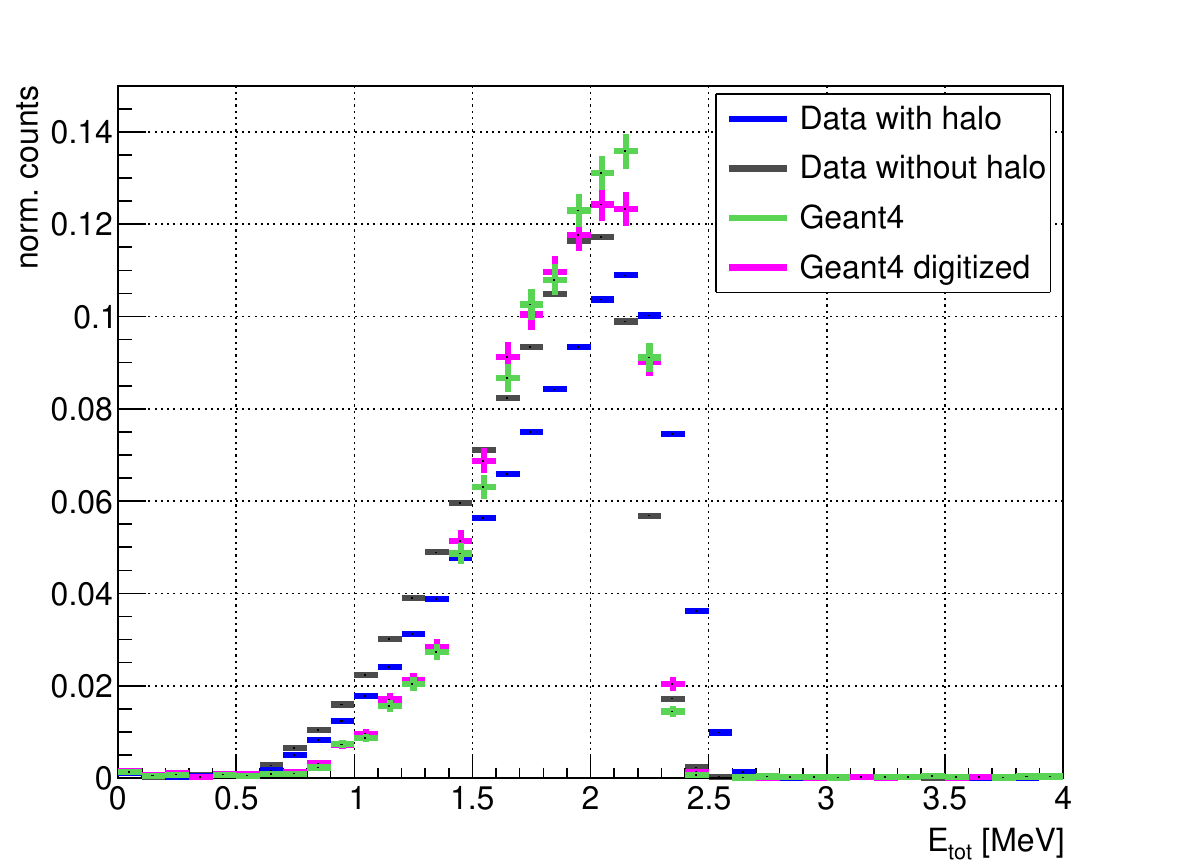}
    \subcaption{}
  \end{minipage}
  \caption{Comparison between data and Geant4 simulations before and after digitisation for the deposited energy for a) cosmic rays and b) $\alpha$ particles from a $\mathrm{^{241}Am}$ radioactive source. For the $\mathrm{^{241}Am}$ the deposited energy before halo removal is also shown for comparison.}
\label{fig:cosmics_alphas_digi}
\end{figure*}

The Timepix3 simulated data were digitised using the $\mathrm{AllPix^{2}}$ modular simulation framework, applying the same bias voltage and charge threshold as during the actual measurements~\cite{allpix, giovanni_thesis}. The clusters generated from MIPs in the cosmic rays simulations reproduced the morphology and the total cluster energy of the measured ones, as observed in Fig.~\ref{fig:cosmics_alphas_digi}a. On the other hand, the digitisation could not consistently reproduce the measured signal from HIPs, due to features of the Timepix3 ASIC, such as the halo, plasma and the volcano effect~\cite{timepix3_heavy_ions_spectro, tpx3_plasma, tpx_pattern_recognition, tpx_stuart_george_thesis, helga_thesis}. These effects arise from substantial energy deposits, and simulating them accurately is challenging because HIPs do not always trigger each of them. The halo (or "skirt"), indicating induced low energy signal in the neighboring pixels is consistently present. However, when heavier or more energetic ions deposit a significant amount of charge over short distances, two additional phenomena may occur: the plasma effect, which extends the charge collection time and consequently increases the track width, or the volcano effect, where pixels with energy deposits exceeding $\sim$500~keV register significantly lower values. The measured cluster energy is then notably lower than the actual deposited energy in the silicon sensor. In addition, the stochastic nature, in particular of the volcano effect, makes their qualitative and quantitative replication in the digitisation unfeasible. Therefore, these effects had to be individually addressed in the data analysis.

The halo pixels, which are easily identifiable due to the very low energy deposits, were excluded implementing an energy cutoff of 5~keV, based on the pixel energy distribution derived from $\mathrm{^{241}Am}$ data. Fig.~\ref{fig:cosmics_alphas_digi}b shows a good agreement of the energy deposition by $\alpha$ particles from a $\mathrm{^{241}Am}$ source between data (after halo removal) and Geant4 simulation, both before and after digitisation, despite the significantly larger clusters observed in data due to the presence of halo pixels. The small discrepancy is likely attributed to the uncertainties in measuring the distance between the source and the detector during the measurements.

\begin{figure}[htp]
\centering

  {\includegraphics[width=0.45\textwidth]{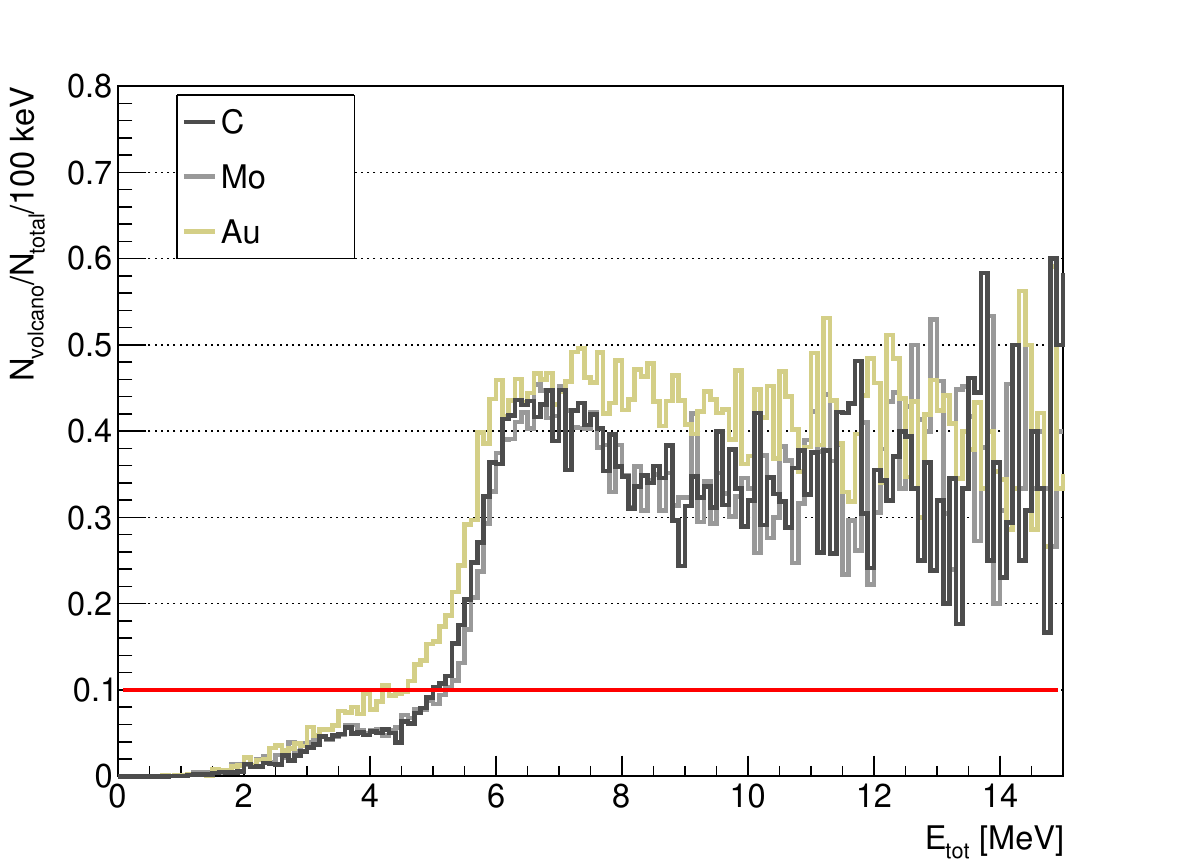}}
  \caption{Fraction of clusters containing volcano pixels, from annihilation measurements with the three different targets, as a function of total energy deposited within the cluster.}
\label{fig:volcano_clusters}
\end{figure}

 Even though the volcano effect manifests in individual pixels, its occurrence in data was examined in relation to the total cluster energy, to reveal the total deposited energy for which this effect has a minimal impact on the measured energy. Such correlation allows to establish an upper limit on the cluster energy that ensures a reliable comparison to simulations. The analysis is presented in Fig.~\ref{fig:volcano_clusters}, where the fraction of clusters with volcano effect is plotted as a function of the deposited energy. Only the clusters produced by HIPs from annihilation in carbon, molybdenum and gold were considered. The figure indicates that up to 3~MeV, the volcano effect has a negligible impact, affecting $<$5\%, while above 5~MeV it is present in $>$10\% of the clusters. An upper limit of 5~MeV was chosen for further comparison of the energy deposited by HIPs to simulations.

\section{\label{sec:Results}Results and discussion}
\subsection{\label{subsec:multiplicities}Multiplicities}
The charged pion multiplicity in antiproton-nucleus annihilation at rest is influenced by the final state interactions of the primary pions with the remaining nucleus. The number of pions that interact with the nucleus is predominantly determined by simple geometric factors, such as the proximity of the annihilation point to the nuclear surface and the size (mass) of the nucleus. The number of surviving, and detectable pions which are not absorbed by the nucleus is related to the primary pions through: 

\begin{eqnarray}
\langle N_{\pi^{\pm}} \rangle = \langle N_{\pi^{\pm}}^{init} \rangle \Bigl(1-\frac{\Omega}{4\pi}\Bigl) 
\end{eqnarray}

where $\langle {N_{\pi^{\pm}}^{init}} \rangle$ = 3.1 is the charged pion multiplicity produced in the initial $\mathrm{\bar{p}}$-nucleon annihilation~\cite{pbar_report_2005}, and $P=(1-\frac{\Omega}{4\pi})$ is the average survival probability for the emitted charged pions~\cite{Cugnon2001_geo_eff}. ${\Omega}$ is the solid angle under which the remaining nucleus is seen from the annihilation point, which is close to the nuclear surface.

\begin{figure}[htp]
  \begin{minipage}{\linewidth}
    \centering
    \includegraphics[width=\textwidth]{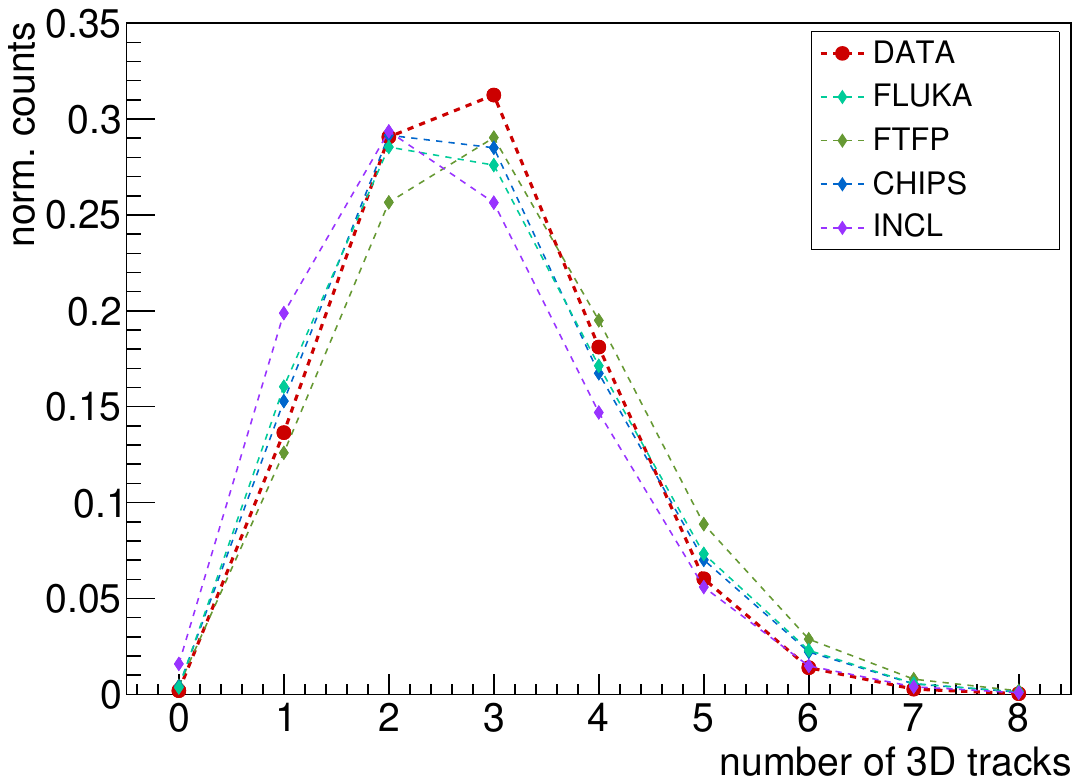}
    \subcaption{}
  \end{minipage}

  \begin{minipage}{\linewidth}
    \centering
    \includegraphics[width=\textwidth]{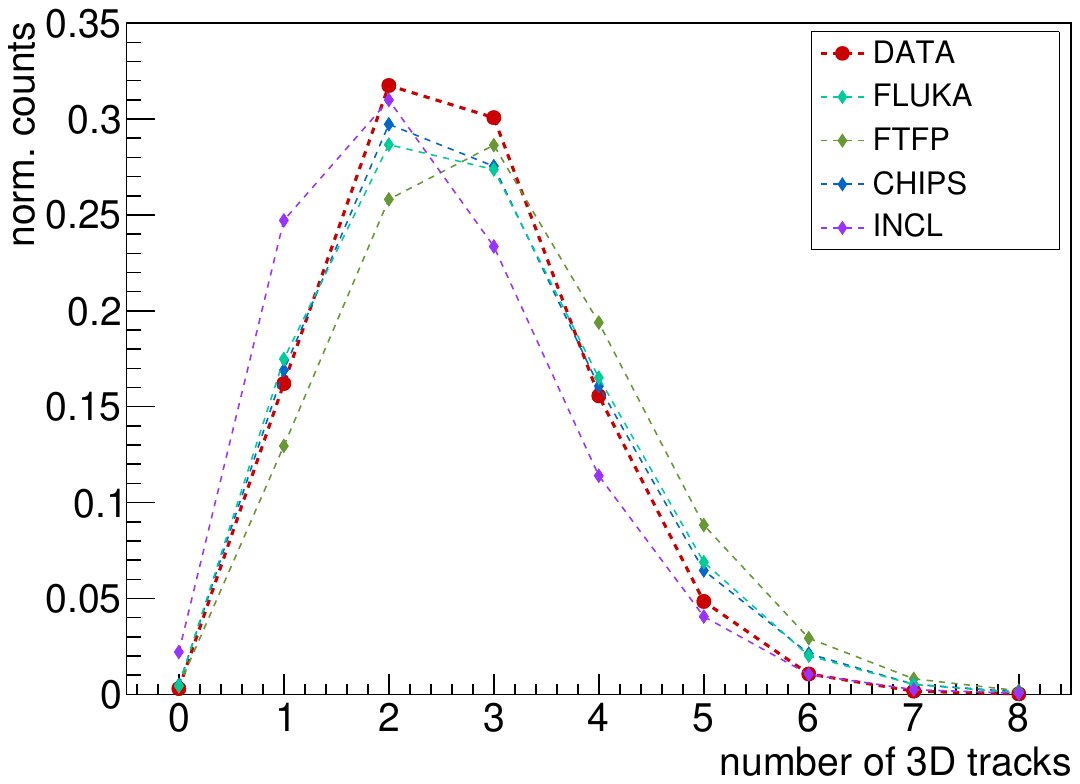}
    \subcaption{}
  \end{minipage}

\begin{minipage}{\linewidth}
    \centering
    \includegraphics[width=\textwidth]{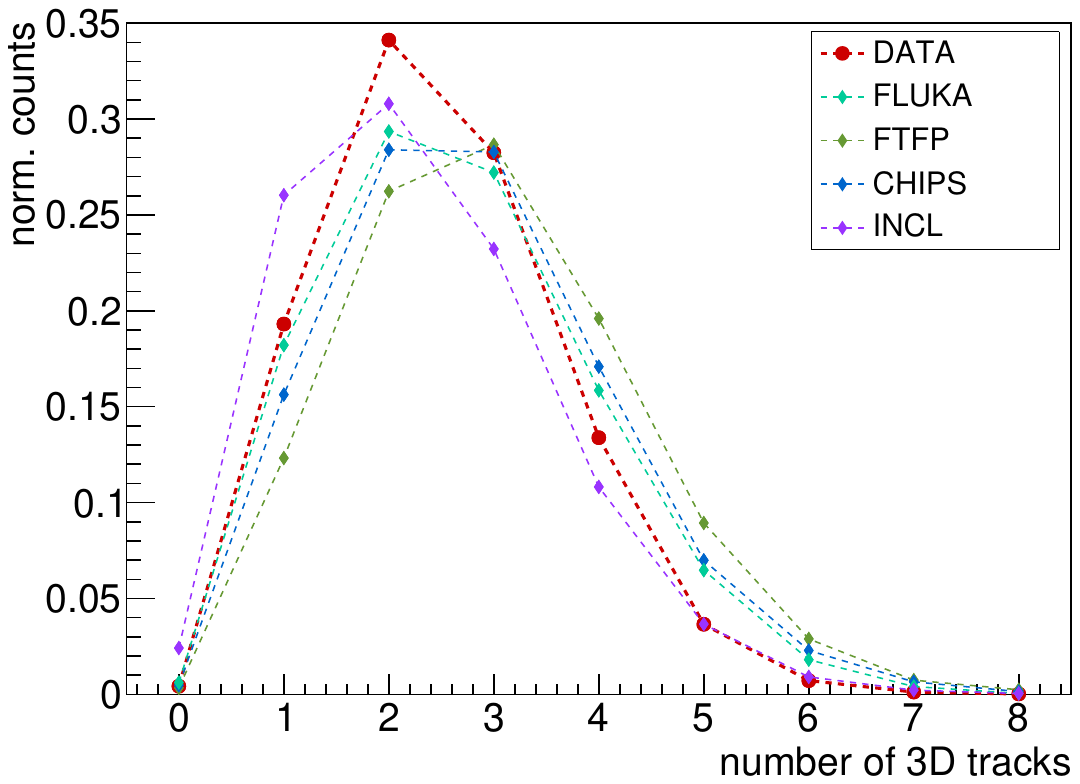}
    \subcaption{}
  \end{minipage}
  
    \caption{Multiplicity distributions of MIPs detected by the hodoscope in measured data and Monte Carlo simulations, for antiproton annihilation at rest with a) carbon, b) molybdenum and c) gold.}
  \label{fig:MIPs_distros}
\end{figure}

Due to the limited geometrical acceptance and efficiency of the hodoscope, only a fraction of the surviving charged pions are measured in this experiment. The distributions of MIP particles presented through the number of reconstructed 3D tracks in the hodoscope are shown in Fig.~\ref{fig:MIPs_distros}. 
No model reproduces accurately the distributions obtained from the measurements for all three targets, with FLUKA, CHIPS and INCL constantly demonstrating more accurate predictions when compared to data, outperforming FTFP. To ensure an authentic comparison with data, the analysis of the simulated data encompassed all events, irrespective of whether antiprotons annihilated in the foil or in its mechanical support ($\sim$7\% of the annihilations), as previously mentioned in Sec.~\ref{sec:intro}. For illustration, excluding antiprotons annihilating on the support structure from the analysis in the Monte Carlo data leads to a marginal change of about 0.2\% of the average MIP multiplicity.

\begin{figure}
\centering
  
  {\includegraphics[width=0.48\textwidth]{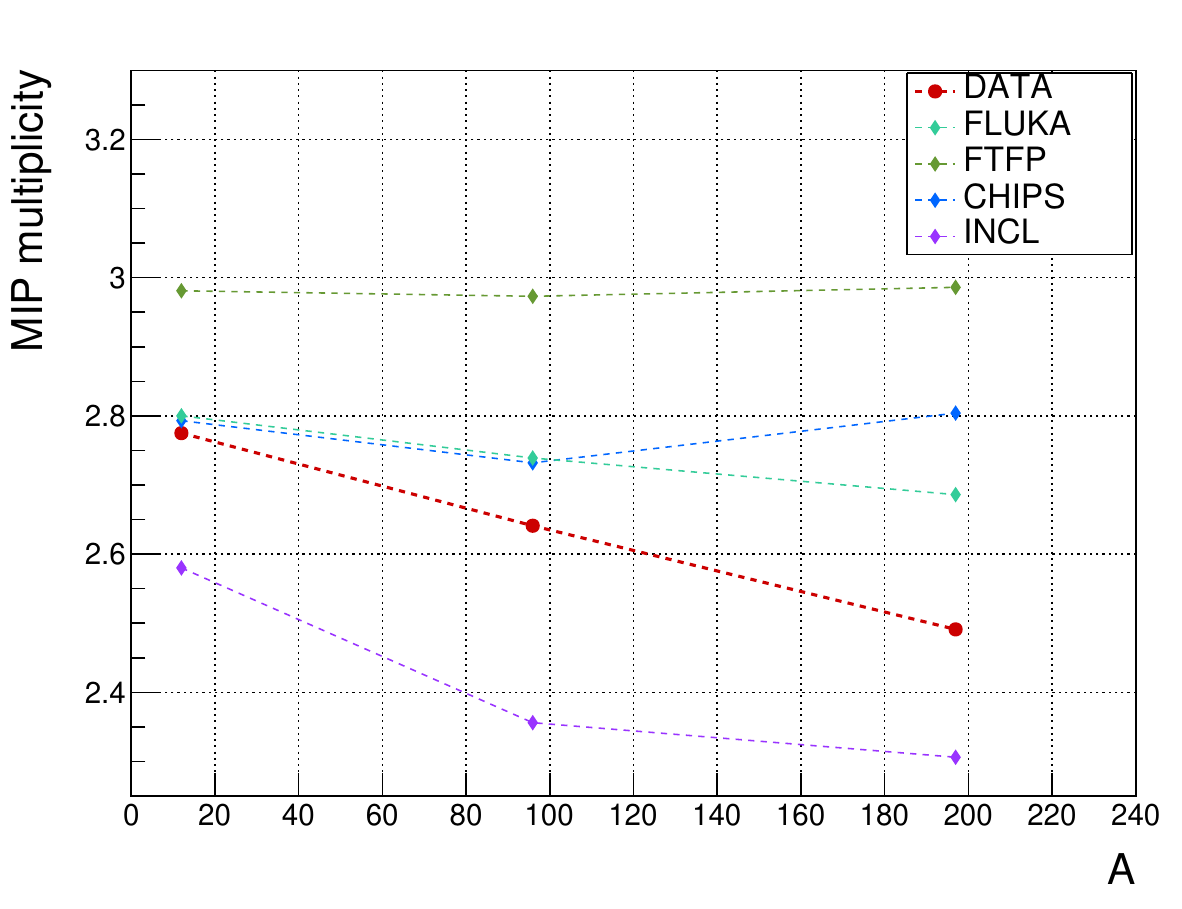}}
  \captionsetup{skip=10pt}
  \caption{Average number of 3D tracks from MIPs in antiproton-nucleus annihilation, detected in the hodoscope, versus atomic mass for carbon, molybdenum and gold targets. Measurements and different Monte Carlo simulation models are compared.} 

\label{fig:MIPs_vs_A}
\end{figure}

A comparison between measurements and simulations of the average MIPs multiplicities detected with the hodoscope are given in Table~\ref{tab:MIPS}, along with the statistical uncertainties. Fig.~\ref{fig:MIPs_vs_A} illustrates these multiplicities as a function of the target atomic mass. The predicted dependence on A is slowly decreasing in FLUKA but is constant for FTFP and CHIPS. INCL constantly underestimates the MIP multiplicities while showing the strongest decrease with A among the models. Previous experimental studies for annihilation of stopped antiprotons indicate a rapid decrease in charged pion multiplicity from A=2 to A$<$80, followed by exponential law, after which it remains nearly constant for higher mass numbers~\cite{Bendiscioli}. In our results, the contribution from particles other than charged pions to the 3D tracks leads rather to a slowly decreasing dependence on A, in line with the FLUKA prediction. However, the disagreement between data and predictions increases with A. For FLUKA, it ranges from $<$1$\%$ for carbon, $4\%$ for molybdenum to 8$\%$ for gold. The CHIPS model predictions are almost as precise as FLUKA for carbon and molybdenum, but are off by $12\%$ for gold. The results obtained with the FTFP model show a discrepancy between about 7$\%$ and 20$\%$ with our experimental data. The new INCL model underestimates the MIP multiplicity between 7\% and 10\%.

\begin{table}
    \centering
    \begin{tabular}{|c|c|c|c|}
    \hline
        \rowcolor{lightgray} & C & Mo & Au\\
         \hline
        \cellcolor{lightgray}Data &~2.775(7)~&~2.642(3)~&~2.491(4)~\\
         \hline
        \cellcolor{lightgray}FLUKA &~2.780(6)~&~2.739(6)~&~2.686(6)~\\
         \hline
        \cellcolor{lightgray}FTFP &~2.981(6)~&~2.973(6)~&~2.987(6)~\\
         \hline
        \cellcolor{lightgray}CHIPS &~2.793(6)~&~2.732(6)~&~2.804(6)~\\
         \hline
        \cellcolor{lightgray}INCL &~2.580(6)~&~2.356(6)~&~2.306(6)~\\
         \hline
    \end{tabular}
    \caption{Average multiplicities of MIPs for antiproton annihilation in carbon, molybdenum and gold, measured with the hodoscope detector.}
    \label{tab:MIPS}
\end{table}

The heavily ionizing particles, including protons and heavier nuclear fragments, were detected with the Timepix3 and tagged with the procedure described in Sec.~\ref{subsec:prongs_in_tpx3}. The low energy antiprotons in this experiment were stopped in the first 100~nm of the target, 
allowing, for example, protons of minimum $\sim$200~keV ($\sim$500~keV) energy to emerge from the 2~${\upmu \mathrm{m}}$ (1~${\upmu \mathrm{m}}$) thick carbon (gold) foil~\cite{icru_Report49}. 
The minimum kinetic energy for a $\mathrm{^{4}He}$ ion to escape the same targets is 800~keV (2~MeV), on the other hand, a $\mathrm{^{12}C}$ ion would need a minimum energy of 5~MeV to break free from the carbon target~\cite{stop_power}. 

\begin{figure}[htp]
  \begin{minipage}{\linewidth}
    \centering
    \includegraphics[width=\textwidth]{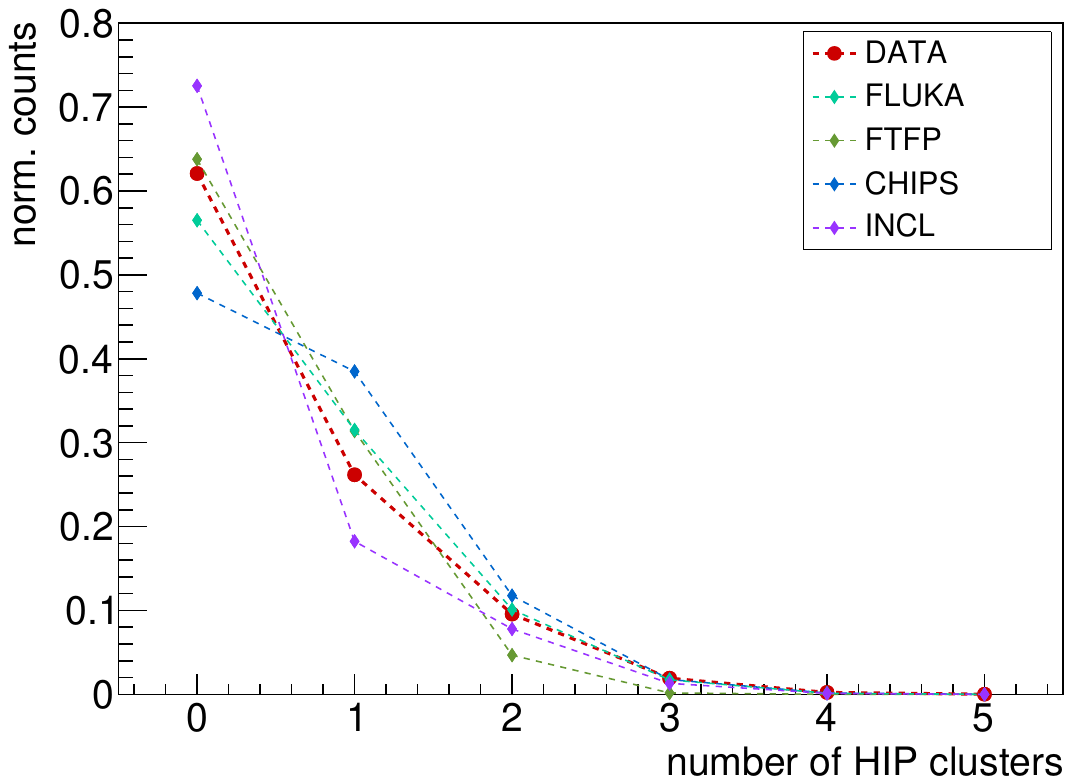}
    \subcaption{}
  \end{minipage}

  \begin{minipage}{\linewidth}
    \centering
    \includegraphics[width=\textwidth]{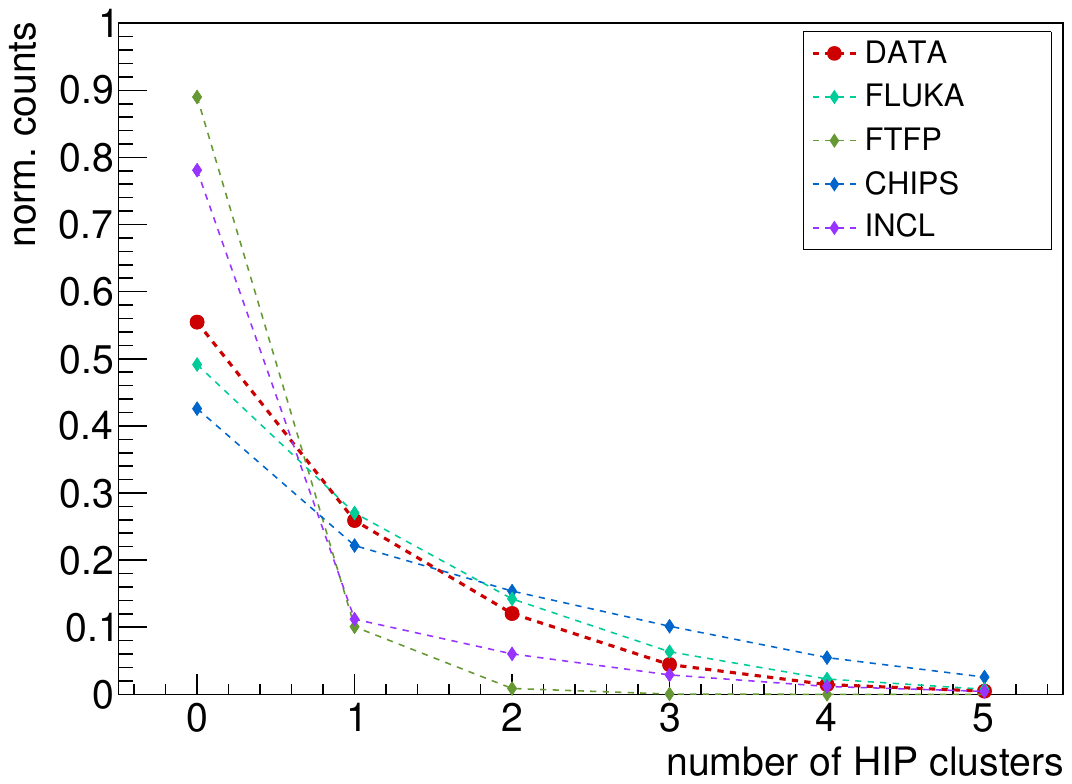}
    \subcaption{}
  \end{minipage}

\begin{minipage}{\linewidth}
    \centering
    \includegraphics[width=\textwidth]{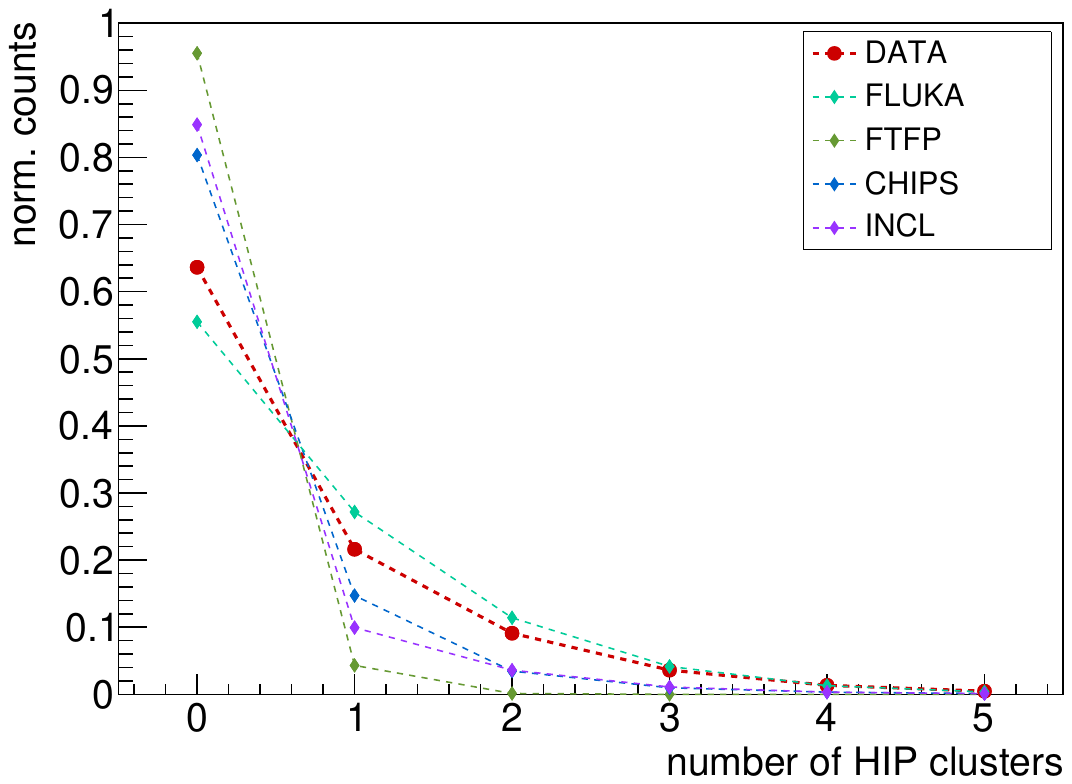}
    \subcaption{}
  \end{minipage}
\caption{Multiplicity distributions of HIPs detected by the Timepix3 quad in measured data and Monte Carlo simulations, for antiproton annihilation in a) carbon, b) molybdenum and c) gold.} 
\label{fig:HIPs_distros}
\end{figure}

In Fig.~\ref{fig:HIPs_distros} the multiplicity distributions of HIPs are compared between data and Monte Carlo simulations for the three target foils. 
Each model is most successful in describing the HIP multiplicity for carbon, the lightest of the three nuclei in this work. FLUKA stands out as the model that provides the best description of the HIPs multiplicity across all three targets, followed by CHIPS. In contrast, FTFP exhibits generally poor agreement with the measured data, and this disparity intensifies with increasing atomic mass.

\begin{table}[h]
    \centering
    \begin{tabular}{|c|c|c|c|}
    \hline
         \rowcolor{lightgray}& C & Mo & Au\\
         \hline
        \cellcolor{lightgray}Data &~0.521(3)~&~0.730(5)~&~0.598(4)~\\
         \hline
        \cellcolor{lightgray}FLUKA &~0.575(3)~&~0.88(1)~&~0.701(4)~\\
         \hline
        \cellcolor{lightgray}FTFP &~0.412(3)~&~0.120(2)~&~0.047(1)~\\
         \hline
        \cellcolor{lightgray}CHIPS &~0.679(3)~&~1.290(7)~&~0.268(3)~\\
         \hline
        \cellcolor{lightgray}INCL &~0.383(8)~&~0.399(8)~&~0.225(3)~\\
         \hline
    \end{tabular}
    \caption{Average multiplicities of HIPs for antiproton annihilation in carbon, molybdenum and gold, measured with the Timepix3 quad detector.}
    \label{tab:HIPS}
\end{table}

For quantitative comparison, Table~\ref{tab:HIPS} presents the average multiplicities for HIPs in the three target foils. The prediction by FLUKA for $\mathrm{\bar{p}}$-C annihilation is 10\% lower than the measured value, while CHIPS and INCL overestimate and underestimate it by 30\%, respectively. FLUKA and CHIPS overestimate the average number of detected HIPs from $\mathrm{\bar{p}}$-Mo annihilation by 20\% and 77\% respectively, while INCL underestimates it by 45\%. For $\mathrm{\bar{p}}$-Au, FLUKA agrees within 17\%, whereas CHIPS and INCL underestimate the data by more than 50\%. For annihilation in carbon, FTFP disagrees with data by $\sim$20\%, while for molybdenum and gold it underestimates the HIP multiplicity by a factor of 6 and 12, respectively.

\begin{figure*}[htp]
\centering
  \begin{minipage}{0.43\textwidth}
    \centering
    \includegraphics[width=\linewidth]{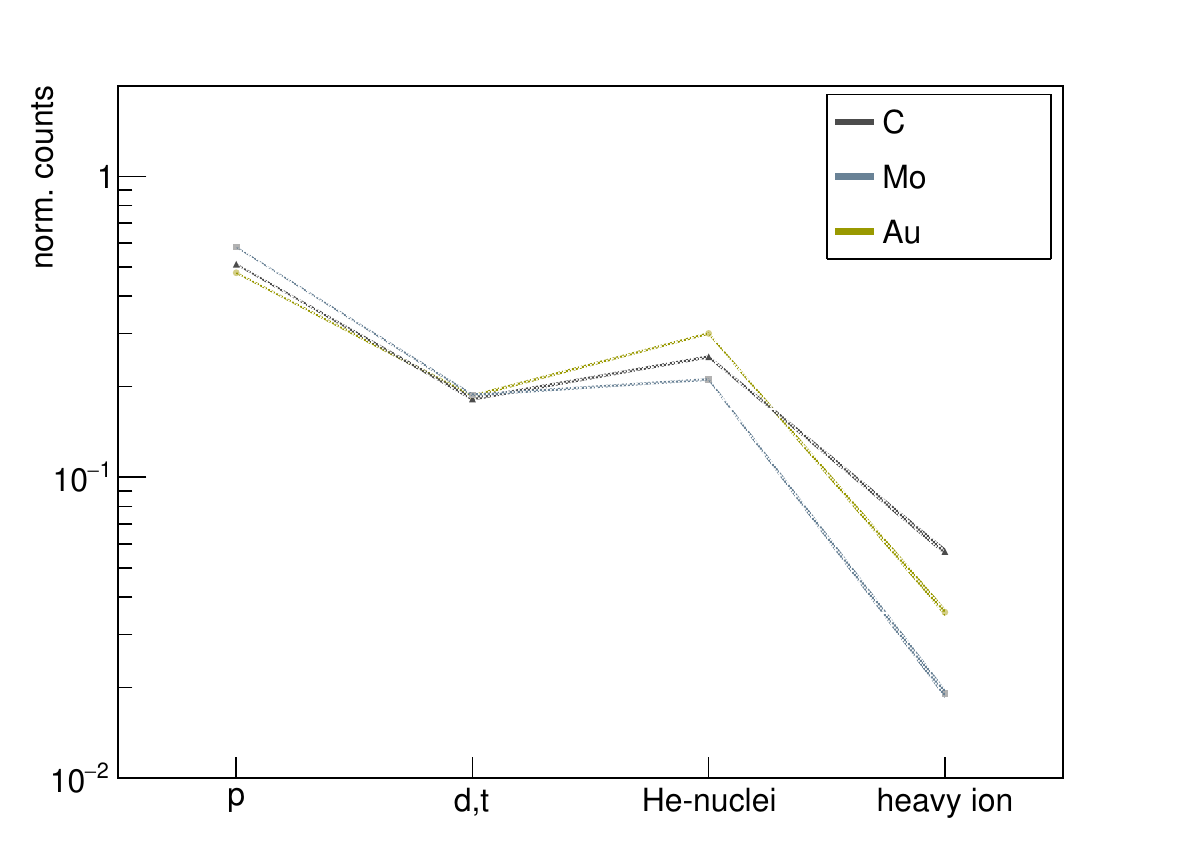}
    \subcaption{}
  \end{minipage}%
  \hspace{0.1\textwidth}
  \begin{minipage}{0.43\textwidth}
    \centering
    \includegraphics[width=\linewidth]{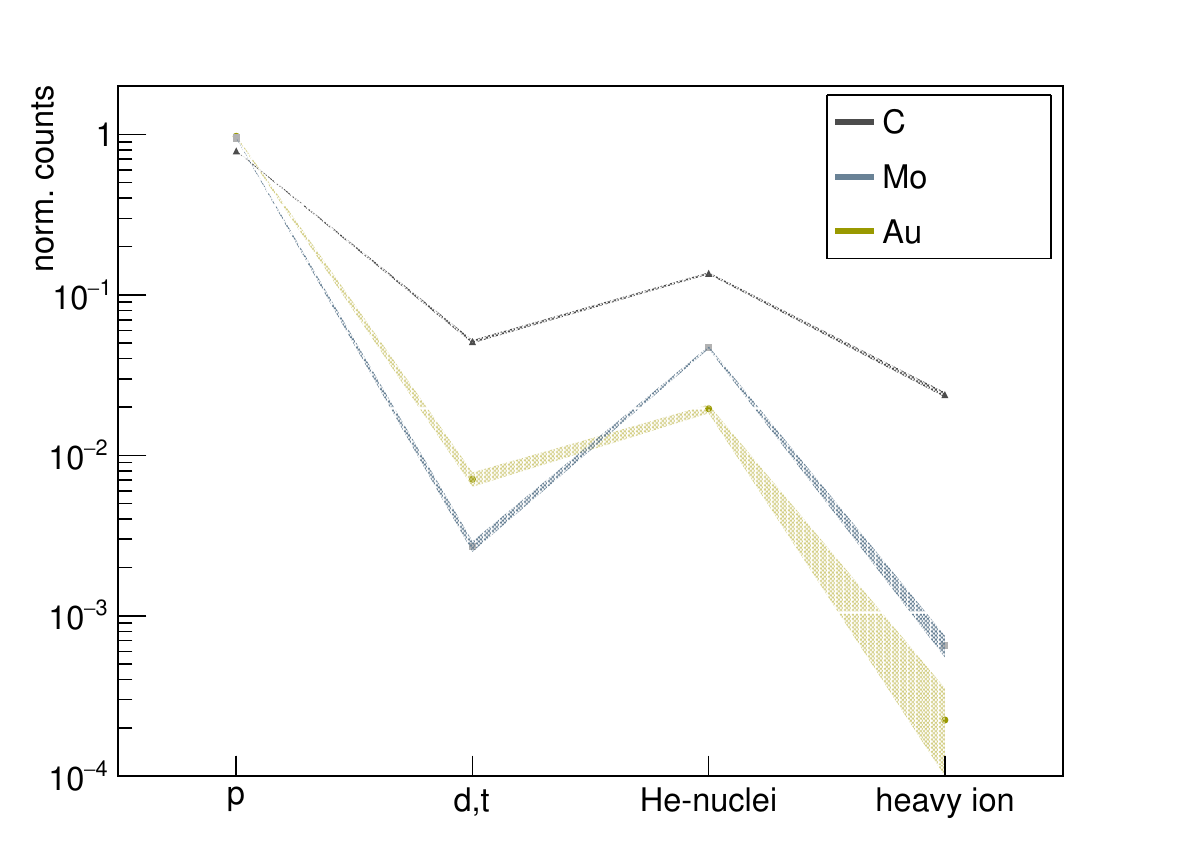}
    \subcaption{}
  \end{minipage}
\caption{Breakdown of the different HIPs detected in the Timepix3 for the two models closest to the measured data, a) FLUKA and b) CHIPS. The linear spline is added to guide the eye, but it also displays the associated error values.} 
\label{fig:sims_HIPs_detail}
\end{figure*}

Fig.~\ref{fig:sims_HIPs_detail} provides a breakdown of the various HIPs generated in $\mathrm{\bar{p}}$ annihilation and detected by the Timepix3, focusing on FLUKA and CHIPS, the two models that most accurately describe the average multiplicity. Protons are dominant nuclear fragments in both models. In FLUKA, about 50\% of HIPs for $\mathrm{\bar{p}}$-C and $\mathrm{\bar{p}}$-Au annihilation are protons, increasing to around 60\% for $\mathrm{\bar{p}}$-Mo. In CHIPS, protons account for nearly 80\% of all HIPs in $\mathrm{\bar{p}}$-C annihilation and exceed 95\% for Mo and Au. While both models feature fragments with Z=2 as the second most represented heavy prongs, they differ by larger factors for Z$>$2. In FLUKA, Z$>$2 fragments constitute between 2\% and 6\% of the total HIPs, while in CHIPS they are similar for $\mathrm{\bar{p}}$-C annihilation ($\sim$2.5\%), but drop to $<$0.1\% for molybdenum and gold.

\subsection{\label{subsec:energy}Energy deposit from heavy prongs}
\begin{figure}[htp]
  \begin{minipage}{\linewidth}
    \centering
    \includegraphics[width=\textwidth]{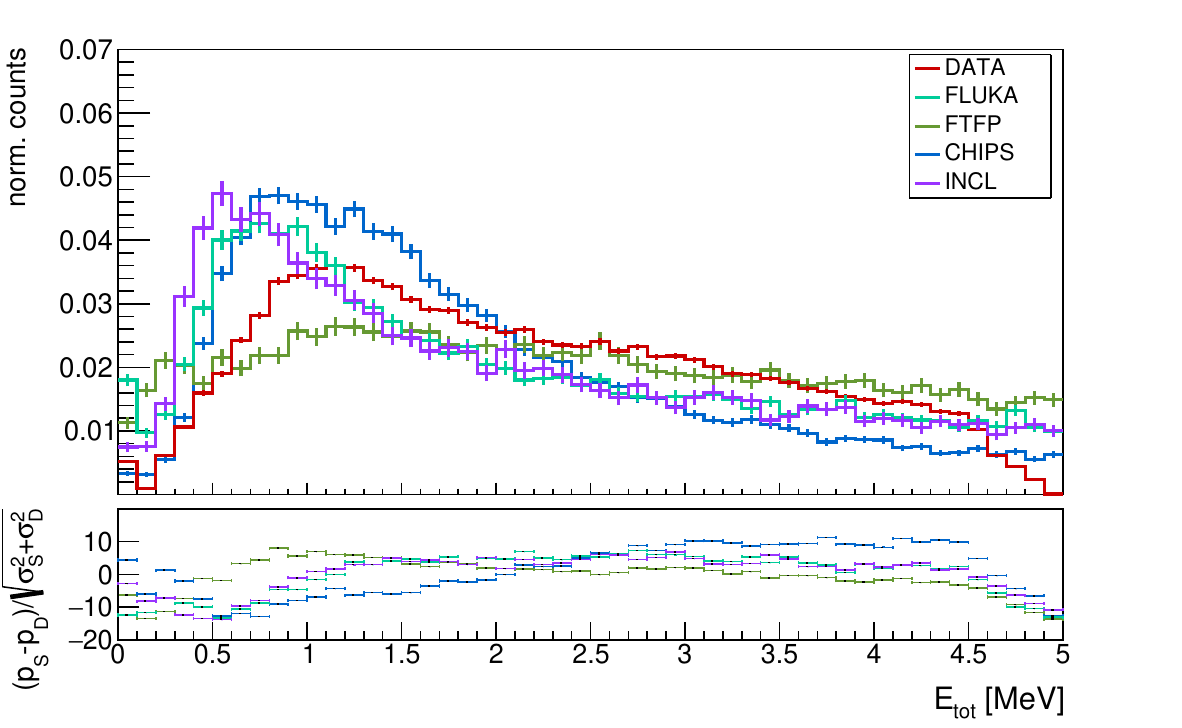}
    \subcaption{}
  \end{minipage}

  \begin{minipage}{\linewidth}
    \centering
    \includegraphics[width=\textwidth]{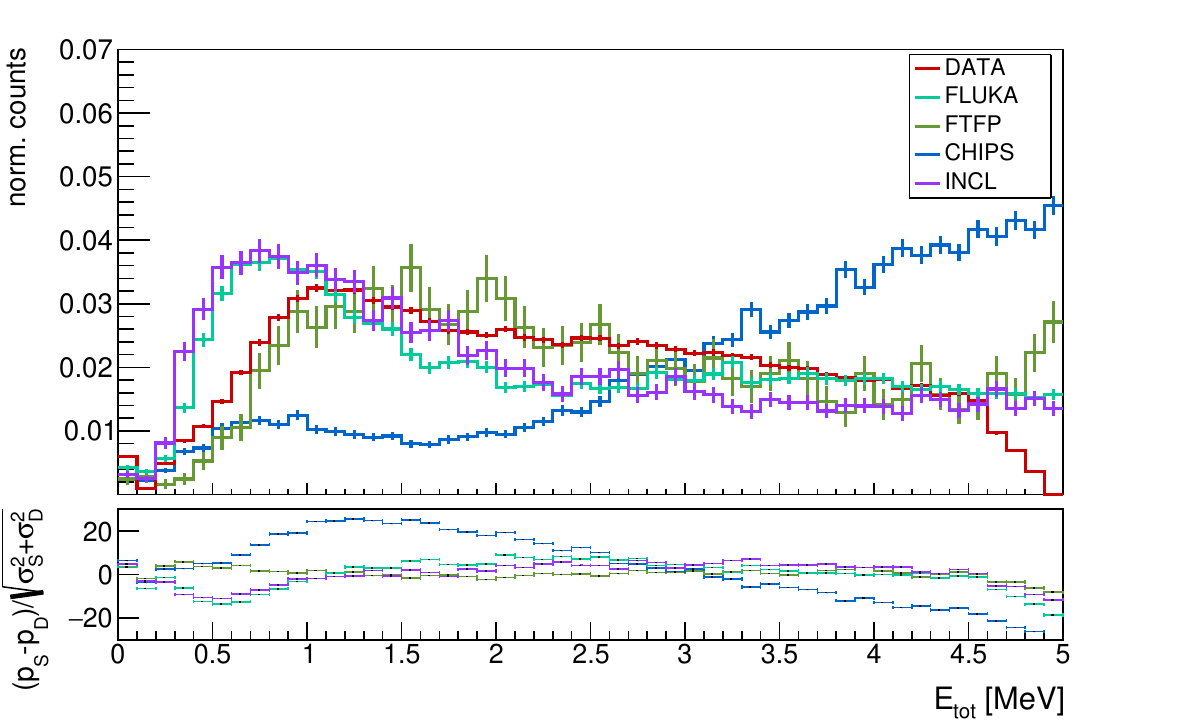}
    \subcaption{}
  \end{minipage}

\begin{minipage}{\linewidth}
    \centering
    \includegraphics[width=\textwidth]{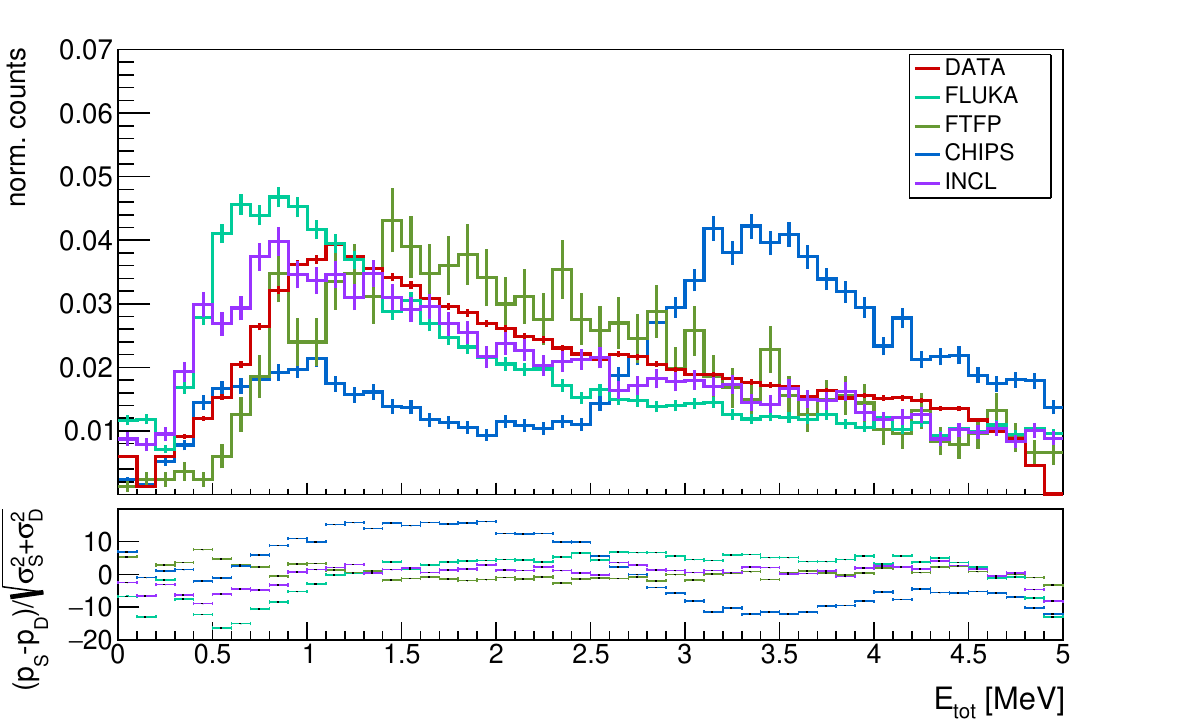}
    \subcaption{}
  \end{minipage}
  
     \caption{Energy deposited by each HIP in the Timepix3 quad for data and Monte Carlo simulations, for antiproton annihilation at rest with a) carbon, b) molybdenum and c) gold. The lower plots show the difference in counts between data and the models in 100~keV bins. Each histogram is normalized to the total number of counts in the range between 0 and 5~MeV.} 
\label{fig:HIPs_Etot}
\end{figure}

The energy deposit from a HIP inside the silicon sensor of the Timepix3 quad detector depends on the type of particle, its kinetic energy as well as the angle under which the particle hits the detector (Sec.~\ref{subsec:detection}). The comparison between data and Monte Carlo models encompasses all HIPs without distinction, and is shown in Fig.~\ref{fig:HIPs_Etot}. The bottom part of the plots shows the difference between data and models in terms of the standard deviation units. As detailed in Sec.~\ref{sec:MC_sims}, the volcano effect, an artifact of the Timepix3 ASIC, constrains the comparison up to 5~MeV.

In $\mathrm{\bar{p}}$A annihilation at rest, protons are emitted through two different processes: evaporation and direct emission. Evaporated protons typically possess energies below $\sim$30~MeV, peaking at $\sim$15~MeV according to ref.~\cite{spectra_pbar}. This energy corresponds to an average deposited energy of $\sim$3~MeV when protons cross the Timepix3 detector perpendicularly, and to $\sim$6.7~MeV when impinging on the detector at the maximum possible angle (67$\degree$) from its normal. 

In our data, energy distributions for HIPs show a broad peak around 1~MeV, which, for illustration, is the average deposited energy for protons with approximately $\sim$60~MeV ($>$100~MeV) traveling perpendicularly (at 67$\degree$ from its normal) through the detector. Protons, deuterons and tritons with energies up to about 10~MeV, 15~MeV, and 22~MeV respectively, deposit more than 5~MeV in the 500~${\upmu \mathrm{m}}$ thick silicon sensor~\cite{stop_power_protons, mimotera}, and are thus excluded form this comparison. $\mathrm{^{3}He}$ and $\mathrm{^{4}He}$ with energies below 30~MeV are fully stopped inside the Timepix3, releasing all of their energy.

Despite the quantitative underestimation of HIP production by FTFP, the energy deposits predicted by this model show the smallest deviation from data for annihilations across all three nuclei. FLUKA exhibits a satisfactory agreement with measurements, with a consistent shift of the maximum of the distribution towards lower energies. This behavior is also characteristic of INCL, whose distributions closely follow those of FLUKA. CHIPS emerges as the model with the least accurate predictions for the deposited energy from HIPs, particularly as the atomic mass of the nucleus increases. 

\section{Conclusions} \label{sec:conclusions}
In this study we measured antiproton-nucleus annihilation at rest in thin targets, with thicknesses of 1~${\upmu \mathrm{m}}$ or 2~${\upmu \mathrm{m}}$, using a slowly extracted antiproton beam of 150~eV. 

To validate Monte Carlo simulation models in the widely-used packages Geant4 and FLUKA, we measured charged particles emitted from the annihilation of antiprotons in carbon, molybdenum, and gold targets. The comparison showed that overall, FLUKA demonstrated the best performance in describing the production of MIPs and HIPs, two quantities tightly bound to the final state interactions between initially produced mesons and residual nuclei. FLUKA's predictions agree with experimental data within about 20\% relative precision, which was confirmed also in ref.~\cite{emulsion2}. The discrepancy between the measured multiplicity of HIPs and FLUKA for Au in our study (15\%) is larger than what was reported in their study (where they were in agreement). A possible explanation for this difference is the variation in target thickness--1~${\upmu \mathrm{m}}$ in our case versus 10~${\upmu \mathrm{m}}$ in theirs. The thinner target allows to detect HIPs over a broader energy range, including protons with energies $<$2~MeV and alpha particles between 2~MeV and 10~MeV, which cannot leave the 10~${\upmu \mathrm{m}}$ target. For heavier ions, this energy range is even wider; for example, for carbon ions, it extends from 5~MeV to 50~MeV. Therefore, the differing levels of agreement between FLUKA and experimental data are likely due to FLUKA overestimating their production.

Among the Geant4 models, CHIPS and INCL show similar performance, outperforming FTFP in reproducing fragment multiplicities. However, CHIPS has been discontinued in the latest versions of Geant4 and is no longer maintained. The primary drawback of FTFP is the underestimation of HIP production by large factors, ranging up to 12. Despite the low rate of produced HIPs, its predictions of the deposited energy of heavy prongs are the most accurate among the examined models. 

Annihilation data at rest using new technologies are needed to test the annihilation mechanisms~\cite{Richard2020}. Future measurements aiming to study the final state interactions will require a more detailed identification of the types of outgoing heavily ionizing particles, for which limited data currently exists. Measuring annihilations in nuclei containing few nucleons, such as Be or C, with this approach can efficiently identify particular FSIs of the annihilation mesons and nucleons. This will be considered in our next project, involving a detailed, systematic study of $\mathrm{\bar{p}}$A annihilation at rest, covering approximately 15 different nuclei. It will be conducted at the AD/ELENA facility at CERN, using the newly developed, dedicated beamline for slow extracted antiprotons at the ASACUSA experiment. The total multiplicity in nearly 4$\pi$ solid angle for charged prongs of various types will be measured, along with their kinetic energy and angular distribution. The project will provide an in-depth investigation of the FSIs, which evolve with the atomic number, enhancing our understanding of antiproton-nucleus interactions at low energies. Moreover, the acquired dataset will be used for detailed validation of the current, but also future models describing antiproton-nucleus annihilation at rest.

\backmatter

\bmhead{Acknowledgements}

This work is supported by JSPS KAKENHI Fostering Joint International Research No.~B~19KK0075, No.~A~20KK0305, Grant-in-Aid for Scientific Research No.~B~20H01930, No.~A~20H00150; the Austrian Science Fund (FWF) Grant Nos. P~32468, W1252-N27; Special Research Projects for Basic Science of RIKEN; Universit{\`a} di Brescia and Istituto Nazionale di Fisica Nucleare; the European Union’s Horizon 2020 research and innovation program under the Marie Sk{\l}odowska-Curie Grant Agreement No.~721559; the Research Grants Program of the Royal Society and the Foundation for Dutch Scientific Research Institutes. The authors express their gratitude to the Medipix Collaboration at CERN for providing the Timepix3 ASICs for the measurements, and to Alberto Ribon from the Geant4 group at CERN for the fruitful discussions and advises.

This research was funded in whole or in part by the Austrian Science Fund (FWF) P~34438. For open access purposes, the author has applied a CC~BY public copyright license to any author accepted manuscript version arising from this submission.

\newpage
\bibliography{sn-article}

\end{document}